\def\mypar{\addvspace{\topsep}\noindent}
\long\def\boxpar#1{\mypar{\fboxsep=6pt\framebox[\linewidth]
  {\begin{minipage}{11.5cm}#1\end{minipage}}}}

\def\ra{\rightarrow}
\def\Ra{\Rightarrow}
\def\bs{\backslash}
\def\bbs{\bs\!\bs}
\def\:{\!:}
\def\qmid{\:}
\def\eeq{\ =\ }
\def\lor{\,\vee\,}
\def\land{\,\wedge\,}
\def\bools{{\mathbb B}}
\def\reals{{\mathbb{R}}}
\def\nats{{\mathbb{N}}}
\def\ints{{\mathbb{Z}}}
\def\two{{\bf 2}}
\def\ext#1{\ulcorner #1\urcorner}
\def\lsem{[\![}
\def\rsem{]\!]}
\def\sem#1{\lsem #1\rsem}
\def\set#1{\{#1\}}
\def\true{{\it true}}

\def\FOR{{\bf for}}
\def\FOREACH{{\bf for each}}
\def\DO{{\bf do}}
\def\UNTIL{{\bf until}}
\def\IF{{\bf if}}
\def\THEN{{\bf then}}
\def\RETURN{{\bf return}}
\def\END{{\bf end}}
\def\EN{\exists\!\bigcirc}
\def\AN{\forall\!\bigcirc}
\def\ED{\exists\Diamond}
\def\EDB#1{\ED_{#1}\,}
\def\AB{\forall\Box}
\def\Next{\bigcirc}
\def\ctl{\mbox{\sc Ctl}}
\def\ltl{\mbox{\sc Ltl}}
\def\ctlstar{\ctl^*}
\def\EB{\exists\mbox{\sc B\"uchi}}
\def\EA{\exists\mbox{\sc A}} 
\def\Buchi{\mbox{\sc B\"uchi}}
\def\U{{\cal U}}
\def\W{{\cal W}}
\def\bfx{\mathbf{x}}

\def\q{\qquad}
\def\qq{\q\q}
\def\qqq{\quad\qq\qq}
\def\qqqq{\qqq\q}

\def\xsem#1{[#1]}
\def\quot#1#2{#1\!/_{#2}}
\def\Sim{{\it Sim}}
\def\dual#1{\overline{#1}}
\def\sts{{\cal S}}
\def\stt{{\cal R}}

\def\states{Q}
\def\regs{R}
\def\obs{P}
\def\trans{\delta}
\def\Pre{{\it Pre}}
\def\And{{\it And}}
\def\Diff{{\it Diff}}
\def\Empty{{\it Empty}}
\def\dEmpty{\dual{\Empty}}
\def\dPre{\dual{\Pre}}
\def\dAnd{\dual{\And}}
\def\dDiff{\dual{\Diff}}
\def\Member{{\it Member}}
\def\state{u}
\def\astate{v}
\def\ob{p}
\def\aob{q}
\def\reg{\sigma}
\def\areg{\tau}
\def\xreg{\rho}
\def\env{E}
\def\xenv{{\cal E}}
\def\log{L}
\def\mlog{L^\mu}
\def\tlog{L^\Diamond}
\def\trace{\pi}
\def\consts{\Pi}
\def\var{x}
\def\vars{X}
\def\hyb{H}
\def\upd{{\it update}}

\def\inv{{\it inv}}
\def\flow{{\it flow}}
\def\hytech{{\sc HyTech}}

\def\lrsem{[\!]}
\newenvironment{cmtab}
  {\begin{center}$\arraycolsep=0pt\begin{array}}{\end{array}$\end{center}}
\newenvironment{chtab}
  {\arraycolsep=0pt\begin{array}{@{\quad\lrsem\ }l@{\ \ra\ }l}}{\end{array}}

\documentclass{llncs}
\usepackage{amssymb,amscd,latexsym,epic,eepic}
\usepackage{fullpage}

\begin{document}

\title{A Classification of Symbolic Transition Systems\thanks{
  This research was supported in part by 
  the DARPA (NASA) grant NAG2-1214,
  the DARPA (Wright-Patterson AFB) grant F33615-C-98-3614,
  the MARCO grant 98-DT-660,
  the ARO MURI grant DAAH-04-96-1-0341, 
  the NSF CAREER award CCR-9501708, 
  and the Belgian National Fund for Scientific Research (FNRS).}}
\author{Thomas~A.~Henzinger$^1$ \qquad Rupak~Majumdar$^1$ \qquad Jean-Fran\c{c}ois Raskin$^2$}
\institute{
  $^1$Department of Electrical Engineering and Computer Sciences\\
  University of California at Berkeley, CA 94720-1770, USA\\
  $^2$D\'epartement d'Informatique, Facult\'e des Sciences\\
  Universit\'e Libre de Bruxelles, Belgium\\
\email{\{tah,rupak,jfr\}@eecs.berkeley.edu}
}
\date{}
\maketitle

\begin{abstract}
We define five increasingly comprehensive classes of infinite-state systems, 
called {\sf STS1--5}, whose state spaces have finitary structure.
For four of these classes, we provide examples from hybrid systems.

\medskip\noindent
{\sf STS1}
These are the systems with finite {\em bisimilarity\/} quotients.
They can be analyzed symbolically by 
(1)~iterating the predecessor and boolean operations starting from a finite 
set of observable state sets, and 
(2)~terminating when no new state sets are generated.
This enables model checking of the $\mu$-calculus.

\medskip\noindent
{\sf STS2}
These are the systems with finite {\em similarity\/} quotients.
They can be analyzed symbolically by iterating the predecessor and positive 
boolean operations.
This enables model checking of the existential and universal fragments of the
$\mu$-calculus.

\medskip\noindent
{\sf STS3}
These are the systems with finite {\em trace-equivalence\/} quotients.
They can be analyzed symbolically by iterating the predecessor operation and 
a restricted form of positive boolean operations 
(intersection is restricted to intersection with observables).
This enables model checking of linear temporal logic.

\medskip\noindent
{\sf STS4}
These are the systems with finite {\em distance-equivalence\/} quotients
(two states are equivalent if for every distance~$d$, the same observables 
can be reached in $d$ transitions).
The systems in this class can be analyzed symbolically by iterating the 
predecessor operation and terminating when no new state sets are generated.
This enables model checking of the existential conjunction-free and universal 
disjunction-free fragments of the $\mu$-calculus. 

\medskip\noindent
{\sf STS5}
These are the systems with finite {\em bounded-reachability\/} quotients
(two states are equivalent if for every distance~$d$, the same observables 
can be reached in $d$ or fewer transitions).
The systems in this class can be analyzed symbolically by iterating the 
predecessor operation and terminating when no new states are encountered.
This enables model checking of reachability properties.
\end{abstract}

\setcounter{section}{-1}
\section{Introduction}

To explore the state space of an infinite-state transition system, it is
often convenient to compute on a data type called ``region,'' whose members 
represent (possibly infinite) sets of states.  
Regions might be implemented, for example, as constraints on the integers or 
reals.
We say that a transition system is ``symbolic'' if it comes equipped with an
algebra of regions which permits the effective computation of certain 
operations on regions. 
For model checking, we are particularly interested in boolean operations on 
regions as well as the predecessor operation, which, given a target region, 
computes the region of all states with successors in the target region.
While a region algebra supports individual operations on regions, the
iteration of these operations may generate an infinite number of distinct 
regions.
In this paper, we study restricted classes of symbolic transition systems for
which certain forms of iteration, if terminated after a finite number of
operations, still yield sufficient information for checking interesting,
unbounded temporal properties of the system.

\subsection{Symbolic Transition Systems}

{\bf Definition: Symbolic transition system}
A {\em symbolic transition system\/}
$\sts=(\states,\trans,\regs,\ext{\cdot},\obs)$
consists of 
  a (possibly infinite) set $\states$ of {\em states},
  a (possibly nondeterministic) {\em transition\/} function
    $\trans\:\states\ra\two^\states$ 
    which maps each state to a set of successor states,
  a (possibly infinite) set $\regs$ of {\em regions},
  an {\em extension\/} function
    $\ext{\cdot}\:\regs\ra\two^\states$ 
    which maps each region to a set of contained states,
  and
  a finite set $\obs\subseteq\regs$ of {\em observables},
such that the following six conditions are satisfied:
\begin{enumerate}
\item
  The set $\obs$ of observables covers the state space~$\states$;
  that is, $\bigcup\set{\ext{\ob}\mid\ob\in\obs}=\states$.
  Moreover, for each observable $\ob\in\obs$, there is a
  complementary observable $\dual{\ob}\in\obs$ such that
  $\ext{\dual{\ob}}=\states\setminus\ext{\ob}$.
\item
  For each region $\reg\in\regs$, there is a region $\Pre(\reg)\in\regs$ such
  that 
    $$\ext{\Pre(\reg)}\eeq
      \set{\state\in\states\mid(\exists\astate\in\trans(\state)\qmid
      \astate\in\reg)};$$
  furthermore, the function $\Pre\:\regs\ra\regs$ is computable.
\item
  For each pair $\reg,\areg\in\regs$ of regions, there is a region 
  $\And(\reg,\areg)\in\regs$ such that 
  $\ext{\And(\reg,\areg)}=\ext{\reg}\cap\ext{\areg}$;
  furthermore, the function $\And\:\regs\times\regs\ra\regs$ is computable.
\item
  For each pair $\reg,\areg\in\regs$ of regions, there is a region 
  $\Diff(\reg,\areg)\in\regs$ such that 
  $\ext{\Diff(\reg,\areg)}=\ext{\reg}\bs\ext{\areg}$;
  furthermore, the function $\Diff\:\regs\times\regs\ra\regs$ is computable.
\item
  All emptiness questions about regions can be decided; 
  that is, there is a computable function $\Empty\:\regs\ra\bools$ such that 
  $\Empty(\reg)$ iff $\ext{\reg}=\emptyset$. 
\item
  All membership questions about regions can be decided; 
  that is, there is a computable function 
  $\Member\:\states\times\regs\ra\bools$ such that 
  $\Member(\state,\reg)$ iff $\state\in\ext{\reg}$. 
\end{enumerate}
The tuple $\stt_\sts=(\obs$,$\Pre$,$\And$,$\Diff$,$\Empty)$ is called the 
{\em region algebra\/} of~$\sts$.
\qed

\mypar
{\bf Remark: Duality}
We take an existential view of symbolic transition systems.
The dual, universal view requires
(1)~$\bigcap\set{\ext{\ob}\mid\ob\in\obs}=\emptyset$, 
(2--4)~closure of $\regs$ under computable functions $\dPre$, $\dAnd$, and 
$\dDiff$ such that
  $$\ext{\dPre(\reg)}\eeq
    \set{\state\in\states\mid(\forall\astate\in\trans(\state)\qmid
    \astate\in\reg)},$$
  $\ext{\dAnd(\reg,\areg)}=\ext{\reg}\cup\ext{\areg}$, and
  $\ext{\dDiff(\reg,\areg)}=\states\bs\ext{\Diff(\areg, \reg)}$,
and 
(5)~a computable function $\dEmpty$ for deciding all universality questions 
about regions (that is, $\dEmpty(\reg)$ iff $\ext{\reg}=\states$). 
All results of this paper have an alternative, dual formulation.
\qed

\mypar
{\bf Remark: Abstract Interpreation}
The region algebra of a symbolic transition system may be viewed as 
the collecting semantics (in the sense of abstract 
interpretation~\cite{CousotCousot77}) of the concrete semantics
of the transition system.
In fact, in a symbolic transition system, the semantics is lifted from
individual states to sets of states.
We refer the interested reader to~\cite{CousotCousot77} for more details
about collecting semantics and abstract interpretation.
\qed

\subsection{Example: Polyhedral Hybrid Automata}

A {\em polyhedral hybrid automaton $\hyb$ of dimension~$m$}, for a positive 
integer~$m$, consists of the following components~\cite{RTSS93j}:
\begin{description}
\item[Continuous variables]
  A set $X=\set{x_1,\ldots,x_m}$ of real-valued variables.
  We write $\dot X$ for the set $\set{\dot x_1,\ldots,\dot x_m}$ of dotted
  variables (which represent first derivatives during continuous change),
  and we write $X'$ for the set $\set{x'_1,\ldots,x'_m}$ of primed variables
  (which represent values at the conclusion of discrete change).
  A {\em linear constraint\/} over $X$ is an expression of the form
  $k_0\sim k_1x_1+\cdots+k_mx_m$, where $\sim\,\in\set{<,\le,=,\ge,>}$ and 
  $k_0,\ldots,k_m$ are integer constants.
  A {\em linear predicate\/} over $X$ is a boolean combination of linear 
  constraints over~$X$.
  Let $L^m$ be the set of linear predicates over~$X$.
\item[Discrete locations]
  A finite directed multigraph $(V,E)$.
  The vertices in $V$ are called {\em locations\/};
  the edges in $E$ are called {\em jumps}.
\item[Invariant and flow conditions]
  Two vertex-labeling functions $\inv$ and $\flow$.
  For each location $v\in V$, the invariant condition $\inv(v)$ is a 
  conjunction of linear constraints over~$X$, and the flow condition 
  $\flow(v)$ is a conjunction of linear constraints over~$\dot X$.
  While the automaton control resides in location~$v$, the variables may 
  evolve according to $\flow(v)$ as long as $\inv(v)$ remains true.
\item[Update conditions]
  An edge-labeling function $\upd$.
  For each jump $e\in E$, the update condition $\upd(e)$ is a conjunction of
  linear constraints over $X\cup X'$.
  The predicate $\upd(e)$ relates the possible values of the variables at the
  beginning of the jump (represented by~$X$) and at the conclusion of the 
  jump (represented by~$X'$).
\end{description}
The polyhedral hybrid automaton $\hyb$ is a 
{\em rectangular automaton\/}~\cite{STOC95j} if 
\begin{verse}
  ---all linear constraints that occur in invariant conditions of $\hyb$ have
  the form $x\sim k$, for $x\in X$ and $k\in\ints$;\\
  ---all linear constraints that occur in flow conditions of $\hyb$ have
  the form $\dot x\sim k$, for $x\in X$ and $k\in\ints$;\\
  ---all linear constraints that occur in jump conditions of $\hyb$ have
  the form $x\sim k$ or $x'=x$ or $x'\sim k$, for $x\in X$ and $k\in\ints$;\\
  ---if $e$ is a jump from location $v$ to location~$v'$, and $\upd(e)$ 
  contains the conjunct $x'=x$, then both $\flow(v)$ and $\flow(v')$ contain
  the same constraints on~$\dot x$.
\end{verse}
The rectangular automaton $\hyb$ is a {\em singular automaton\/} if each flow
condition of $\hyb$ has the form
$\dot{x}_1=k_1\wedge\ldots\wedge\dot{x}_m=k_m$.
The singular automaton $\hyb$ is a {\em timed automaton\/}~\cite{AlurDill94} 
if each flow condition of $\hyb$ has the form
$\dot{x}_1=1\wedge\ldots\wedge\dot{x}_m=1$.

\mypar
The polyhedral hybrid automaton $\hyb$ defines the symbolic transition system 
$\sts_\hyb=(\states_\hyb,\trans_\hyb,\regs_\hyb,\ext{\cdot}_\hyb,\obs_\hyb)$
with the following components:
\begin{itemize}
\item
  $\states_\hyb=V\times\reals^m$; 
  that is, every state $(v,\bfx)$ consists of a location $v$
  (the discrete component of the state) and values $\bfx$ for the 
  variables in~$X$ (the continuous component).
\item
  $(v',\bfx')\in\trans_{\hyb}(v,\bfx)$ if either 
  (1) there is a jump $e\in E$ from $v$ to $v'$ such that the closed 
  predicate $\upd(e)[X,X':=\bfx,\bfx']$ is true, or 
  (2)~$v'=v$ and there is a real $\Delta\ge 0$ and a differentiable function 
  $f\:[0,\Delta]\ra\reals^m$ with first derivative $\dot f$ such that 
  $f(0)=\bfx$ and $f(\Delta)=\bfx'$, and for all reals 
  $\varepsilon\in(0,\Delta)$, the closed predicates 
  $\inv(v)[X:=f(\varepsilon)]$ and 
  $\flow(v)[\dot X:=\dot{f}(\varepsilon)]$ are true.
  In case~(2), the function $f$ is called a {\em flow function}.
\item
  $\regs_\hyb=V\times L^m$; 
  that is, every region $(v,\phi)$ consists of a location $v$ 
  (the discrete component of the region) and a linear predicate $\phi$ 
  over~$X$ (the continuous component).
\item
  $\ext{(v,\phi)}_\hyb=\set{(v,\bfx)\mid 
    \bfx\in\reals^m\mbox{ and }\phi[X:=\bfx]\mbox{ is true}}$;
  that is, the extension function maps the continuous component $\phi$ of a 
  region to the values for the variables in $X$ which satisfy the
  predicate~$\phi$. 
  Consequently, the extension of every region consists of a location and a 
  polyhedral subset of~$\reals^m$.
\item 
  $\obs_\hyb=V\times\set{\true}$;
  that is, only the discrete component of a state is observable.
\end{itemize}
It requires some work to see that $\sts_\hyb$ is indeed a symbolic transition
system.
First, notice that the linear predicates over $X$ are closed under all 
boolean operations, and that satisfiability is decidable for the linear 
predicates.
Second, the $\Pre$ operator is computable on~$\regs_\hyb$, because all flow 
functions can be replaced by straight lines~\cite{RTSS93j}.

\subsection{Background Definitions}

The symbolic transition systems are a special case of transition systems.
A {\em transition system\/} 
$\sts=(\states,\trans,\cdot,\ext{\cdot},\obs)$ has the same components as a 
symbolic transition system, except that no regions are specified and the 
extension function is defined only for the observables
(that is, $\ext{\cdot}\:\obs\ra\two^\states$).

\mypar
{\bf State equivalences}
A {\em state equivalence\/} $\cong$ is a family of relations which contains 
for each transition system $\sts$ an equivalence relation $\cong^\sts$ on the
states of~$\sts$.
The $\cong$ {\em equivalence problem\/} for a class {\sf C} of transition 
systems asks, given two states $\state$ and $\astate$ of a transition system
$\sts$ from the class~{\sf C}, whether $\state\cong^\sts\astate$. 
The state equivalence $\cong_a$ is {\em as coarse as\/} the state equivalence
$\cong_b$ if $\state\cong_a^\sts\astate$ implies $\state\cong_b^\sts\astate$
for all transition systems~$\sts$.
The equivalence $\cong_a$ is {\em coarser than\/} $\cong_b$ if $\cong_a$ is 
as coarse as~$\cong_b$, but $\cong_b$ is not as coarse as~$\cong_a$.
Given a transition system $\sts=(\states,\trans,\cdot,\ext{\cdot},\obs)$ and 
a state equivalence~$\cong$, the {\em quotient system\/} is the transition 
system
  $\quot{\sts}{\cong}=(\quot{\states}{\cong},\quot{\delta}{\cong},\cdot,
    \quot{\ext{\cdot}}{\cong},\obs)$
with the following components:
\begin{verse}
  ---the states in $\quot{\sts}{\cong}$ are the equivalence classes 
  of~$\cong_\sts$;\\
  ---$\areg\in\quot{\delta}{\cong}(\reg)$ if there is a state $\state\in\reg$ 
  and a state $\astate\in\areg$ such that $\astate\in\delta(\state)$;\\
  ---$\reg\in\quot{\ext{\ob}}{\cong}$ if there is a state $\state\in\reg$ such 
  that $\state\in\ext{\ob}$.
\end{verse}
The quotient construction is of particular interest to us when it transforms 
an infinite-state system $\sts$ into a finite-state 
system~$\quot{\sts}{\cong}$.

\mypar
{\bf State logics}
A {\em state logic\/} $\log$ is a logic whose formulas are interpreted over
the states of transition systems;
that is, for every $\log$-formula $\varphi$ and every transition
system~$\sts$, there is a set $\sem{\varphi}_\sts$ of states of $\sts$ which
satisfy~$\varphi$.
The $\log$ {\em model-checking problem\/} for a class {\sf C} of transition 
systems asks, given an $\log$-formula $\varphi$ and a state $\state$ of a 
transition system $\sts$ from the class~{\sf C}, whether 
$\state\in\sem{\varphi}_\sts$.
Two formulas $\varphi$ and $\psi$ of state logics are {\em equivalent\/} if
$\sem{\varphi}_\sts=\sem{\psi}_\sts$ for all transition systems~$\sts$.
The state logic $\log_a$ is {\em as expressive as\/} the state logic
$\log_b$ if for every $\log_b$-formula~$\varphi$, there is an
$\log_a$-formula $\psi$ which is equivalent to~$\varphi$.
The logic $\log_a$ is {\em more expressive than\/} $\log_b$ if $\log_a$ is 
as expressive as~$\log_b$, but $\log_b$ is not as expressive as~$\log_a$.
Every state logic $\log$ {\em induces\/} a state equivalence, 
denoted~$\cong_{\log}$:
for all states $\state$ and $\astate$ of a transition system~$\sts$, define
$\state\cong_{\log}^\sts\astate$ if for all $\log$-formulas~$\varphi$, we 
have $\state\in\sem{\varphi}_\sts$ iff $\astate\in\sem{\varphi}_\sts$.
The state logic $\log$ {\em admits abstraction\/} if for every $\log$-formula
$\varphi$ and every transition system~$\sts$, we have
  $\sem{\varphi}_\sts=
    \bigcup\set{\reg\mid\reg\in\sem{\varphi}_{\quot{\sts}{\cong_{\log}}}}$;
that is, a state $\state$ of $\sts$ satisfies an $\log$-formula $\varphi$ iff
the $\cong_{\log}$ equivalence class of $\state$ satisfies $\varphi$ in the 
quotient system.
Consequently, if $\log$ admits abstraction, then every $\log$ model-checking 
question on a transition system $\sts$ can be reduced to an $\log$ 
model-checking question on the induced quotient 
system~$\quot{\sts}{\cong_{\log}}$.
Below, we shall repeatedly prove the $\log$ model-checking problem for a
class {\sf C} to be decidable by observing that for every transition system
$\sts$ from {\sf C}, the quotient system $\quot{\sts}{\cong_{\log}}$ has
finitely many states and can be constructed effectively.

\mypar
{\bf Symbolic semi-algorithms}
A {\em symbolic semi-algorithm\/} takes as input the region algebra 
$\stt_\sts=$ $(\obs$, $\Pre$, $\And$, $\Diff$, $\Empty)$ of a symbolic transition 
system $\sts=(\states,\trans,\regs,\ext{\cdot},\obs)$, and generates regions 
in $\regs$ using the operations $\obs$, $\Pre$, $\And$, $\Diff$, and
$\Empty$. 
Depending on the input~$\sts$, a symbolic semi-algorithm on $\sts$ may or may
not terminate.

\subsection{Preview}

In sections 1--5 of this paper, we shall define five increasingly 
comprehensive classes of symbolic transition systems.
In each case $i\in\set{1,\ldots,5}$, we will proceed in four steps:

\mypar
{\bf 1 Definition: Finite characterization}
  We give a state equivalence $\cong_i$ and define the class {\sf STS}$(i)$ 
  to contain precisely the symbolic transition systems $\sts$ for which the 
  equivalence relation $\cong_i^\sts$ has finite index 
  (i.e., there are finitely many $\cong_i^\sts$ equivalence classes).
  Each state equivalence $\cong_i$ is coarser than its 
  predecessor~$\cong_{i-1}$, which implies that 
  {\sf STS}$(i-1)$ $\subsetneq$ {\sf STS}$(i)$ for $i\in\set{2,\ldots,5}$.

\mypar
{\bf 2 Algorithmics: Symbolic state-space exploration}
  We give a symbolic semi-algorithm that terminates precisely on the symbolic
  transition systems in the class {\sf STS}$(i)$.
  This provides an operational characterization of the class {\sf STS}$(i)$ 
  which is equivalent to the denotational definition of {\sf STS}$(i)$.
  Termination of the semi-algorithm is proved by observing that if given the 
  region algebra of a symbolic transition system $\sts$ as input, then the
  extensions of all regions generated by the semi-algorithm are 
  $\cong_i^\sts$ blocks (i.e., unions of $\cong_i^\sts$ equivalence classes).
  If $\sts$ is in the class {\sf STS}$(i)$, then there are only finitely many
  $\cong_i^\sts$ blocks, and the semi-algorithm terminates upon having 
  constructed a representation of the quotient system~$\quot{\sts}{\cong_i}$.
  The semi-algorithm can therefore be used to decide all $\cong_i$ 
  equivalence questions for the class {\sf STS}$(i)$.

\mypar
{\bf 3 Verification: Decidable properties}
  We give a state logic $\log_i$ which admits abstraction and induces the 
  state equivalence~$\cong_i$.
  Since $\cong_i$ quotients can be constructed effectively, it follows that 
  the $\log_i$ model-checking problem for the class {\sf STS}$(i)$ is 
  decidable.
  However, model-checking algorithms which rely on the explicit construction 
  of quotient systems are usually impractical.
  Hence, we also give a symbolic semi-algorithm that terminates on the 
  symbolic transition systems in the class {\sf STS}$(i)$ and directly 
  decides all $\log_i$ model-checking questions for this class.

\mypar
{\bf 4 Example: Hybrid systems}
  The interesting members of the class {\sf STS}$(i)$ are those with 
  infinitely many states.
  In four out of the five cases, following~\cite{LICS96H}, we provide certain 
  kinds of polyhedral hybrid automata as examples.

\section{Class-1 Symbolic Transition Systems}

Class-1 systems are characterized by finite bisimilarity quotients.
The region algebra of a class-1 system has a finite subalgebra that contains
the observables and is closed under $\Pre$, $\And$, and $\Diff$ operations.
This enables the model checking of all $\mu$-calculus properties.
Infinite-state examples of class-1 systems are provided by the singular
hybrid automata.

\subsection{Finite Characterization: Bisimilarity}

{\bf Definition: Bisimilarity}
Let $\sts=(\states,\trans,\cdot,\ext{\cdot},\obs)$ be a transition system.
A binary relation $\preceq$ on the state space $\states$ is a 
{\em simulation\/} on $\sts$ if $\state\preceq\astate$ implies the following 
two conditions:
\begin{verse}
  1. For each observable $\ob\in\obs$, we have $\state\in\ext{\ob}$ iff 
     $\astate\in\ext{\ob}$.\\
  2. For each state $\state'\in\trans(\state)$, there is a state 
     $\astate'\in\trans(\astate)$ such that $\state'\preceq\astate'$. 
\end{verse}
Two states $\state,\astate\in\states$ are {\em bisimilar}, denoted
$\state\cong_1^\sts\astate$, if there is a symmetric simulation $\preceq$ on 
$\sts$ such that $\state\preceq\astate$. 
The state equivalence $\cong_1$ is called {\em bisimilarity}.
\qed

\mypar
{\bf Definition: Class {\sf STS1}}
A symbolic transition system $\sts$ belongs to the class {\sf STS1} if the 
bisimilarity relation $\cong_1^\sts$ has finite index. 
\qed

\subsection{Symbolic State-space Exploration: Partition Refinement}

The bisimilarity relation of a finite-state system can be computed by 
partition refinement~\cite{KanellakisSmolka90}. 
The symbolic semi-algorithm {\sf Closure1} of Figure~\ref{closure1} applies 
this method to infinite-state systems \cite{BFH90,ICALP95H}.
Suppose that the input given to {\sf Closure1} is the region algebra of a 
symbolic transition system $\sts=(\states,\trans,\regs,\ext{\cdot},\obs)$.
Then each~$T_i$, for $i\ge 0$, is a finite set of regions;
that is, $T_i\subseteq\regs$.
By induction it is easy to check that for all $i\ge 0$, the extension of 
every region in $T_i$ is a $\cong_1^\sts$ block.
Thus, if $\cong_1^\sts$ has finite index, then {\sf Closure1} terminates.
Conversely, suppose that {\sf Closure1} terminates with 
$\ext{T_{i+1}}\subseteq\ext{T_i}$.
{F}rom the definition of bisimilarity it follows that if for each region
$\reg\in T_i$, we have $s\in\ext{\reg}$ iff $t\in\ext{\reg}$, then
$\state\cong_1^\sts\astate$.
This implies that $\cong_1^\sts$ has finite index.

\begin{figure}[t]
\boxpar{
\begin{verse}
  {\bf Symbolic semi-algorithm} {\sf Closure1}\\
  Input: a region algebra $\stt=(\obs,\Pre,\And,\Diff,\Empty)$.\\[5pt]
  \q $T_0$ := $\obs$;\\
  \q \FOR\ $i=0,1,2,\ldots$ \DO\\
  \qq $T_{i+1}$ := $T_i$\\
  \qqq $\cup\ \set{\Pre(\reg)\mid\reg\in T_i}$\\
  \qqq $\cup\ \set{\And(\reg,\areg)\mid\reg,\areg\in T_i}$\\
  \qqq $\cup\ \set{\Diff(\reg,\areg)\mid\reg,\areg\in T_i}$\\
  \qq \UNTIL\ $\ext{T_{i+1}}\subseteq\ext{T_i}$. 
\end{verse}
The termination test $\ext{T_{i+1}}\subseteq\ext{T_i}$, which is shorthand for
$\set{\ext{\reg}\mid\reg\in T_{i+1}}\subseteq
  \set{\ext{\reg}\mid\reg\in T_i}$,
is decided as follows:
for each region $\reg\in T_{i+1}$ check that there is a region $\areg\in T_i$
such that both $\Empty(\Diff(\reg,\areg))$ and $\Empty(\Diff(\areg,\reg))$.
}
\caption{Partition refinement}
\label{closure1}
\end{figure}

\mypar
{\bf Theorem 1A} 
{\it 
For all symbolic transition systems~$\sts$, the symbolic semi-algorithm 
{\sf Closure1} terminates on the region algebra $\stt_\sts$ iff $\sts$ 
belongs to the class {\sf STS1}.
}

\mypar
{\bf Corollary 1A} 
{\it 
The $\cong_1$ (bisimilarity) equivalence problem is decidable for the class 
{\sf STS1} of symbolic transition systems.
}

\subsection{Decidable Properties: Branching Time}

{\bf Definition: $\mu$-calculus}
The formulas of the {\em $\mu$-calculus\/} are generated by the grammar
  \[\varphi\ ::=\
    \ob \mid
    \dual{\ob} \mid
    \var \mid
    \varphi\vee\varphi \mid 
    \varphi\wedge\varphi \mid 
    \EN\varphi \mid 
    \AN\varphi \mid 
    (\mu \var\qmid\varphi) \mid 
    (\nu \var\qmid\varphi),
    \]
for constants $\ob$ from some set~$\consts$, and variables $\var$ from some 
set~$\vars$.
Let $\sts=(\states,\trans,\cdot,\ext{\cdot},\obs)$ be a transition system 
whose observables include all constants; 
that is, $\consts\subseteq\obs$.
Let $\xenv\:\vars\ra\two^\states$ be a mapping from the variables to sets of 
states.
We write $\xenv[\var\mapsto\xreg]$ for the mapping that agrees with $\xenv$ on 
all variables, except that $\var\in\vars$ is mapped to $\xreg\subseteq\states$.
Given $\sts$ and~$\xenv$, every formula $\varphi$ of the $\mu$-calculus 
defines a set $\sem{\varphi}_{\sts,\xenv}\subseteq\states$ of states:
\begin{verse}
  $\sem{\ob}_{\sts,\xenv}\eeq\ext{\ob}$;\\
  $\sem{\dual{\ob}}_{\sts,\xenv}\eeq\states\bs\ext{\ob}$;\\
  $\sem{\var}_{\sts,\xenv}\eeq\xenv(\var)$;\\
  $\sem{\varphi_1{\vee\brace\wedge}\varphi_2}_{\sts,\xenv}\eeq
    \sem{\varphi_1}_{\sts,\xenv}\ {\cup\brace\cap}\
    \sem{\varphi_2}_{\sts,\xenv}$;\\
  $\sem{{\exists\brace\forall}\!\Next\varphi}_{\sts,\xenv}\eeq
    \set{\state\in\states\mid({\exists\brace\forall}\astate\in\trans(\state)
    \qmid\astate\in\sem{\varphi}_{\sts,\xenv})}$;\\
  $\sem{{\mu\brace\nu}\var\qmid\varphi}_{\sts,\xenv}\eeq
    {\cap\brace\cup}\set{\xreg\subseteq\states\mid
    \xreg=\sem{\varphi}_{\sts,\xenv[\var\mapsto\xreg]}}$.
\end{verse}
If we restrict ourselves to the closed formulas of the $\mu$-calculus, then 
we obtain a state logic, denoted~$\mlog_1$:
the state $\state\in\states$ {\em satisfies\/} the $\mlog_1$-formula $\varphi$
if $\state\in\sem{\varphi}_{\sts,\xenv}$ for any variable mapping~$\xenv$;
that is, $\sem{\varphi}_\sts=\sem{\varphi}_{\sts,\xenv}$ for any~$\xenv$.
\qed

\mypar
{\bf Remark: Duality} 
For every $\mlog_1$-formula~$\varphi$, the {\em dual\/} $\mlog_1$-formula 
$\dual{\varphi}$ is obtained by replacing the constructors $\ob$, 
$\dual{\ob}$, $\vee$, $\wedge$, $\EN$, $\AN$, $\mu$, and $\nu$ by 
$\dual{\ob}$, $\ob$, $\wedge$, $\vee$, $\AN$, $\EN$, $\nu$, and~$\mu$, 
respectively.
Then, $\sem{\dual{\varphi}}_\sts=\states\bs\sem{\varphi}_\sts$.
It follows that the answer of the model-checking question for a state 
$\state\in\states$ and an $\mlog_1$-formula $\varphi$ is complementary to the 
answer of the model-checking question for $\state$ and the dual 
formula~$\dual{\varphi}$.
\qed

\mypar
The following facts about the $\mu$-calculus are relevant in our 
context~\cite{CavBook}.
First, $\mlog_1$ admits abstraction, and the state equivalence induced by
$\mlog_1$ is~$\cong_1$ (bisimilarity).
Second, $\mlog_1$ is very expressive;
in particular, $\mlog_1$ is more expressive than the temporal logics 
$\ctlstar$ and $\ctl$, which also induce bisimilarity.
Third, the definition of $\mlog_1$ naturally suggests a model-checking 
method for finite-state systems, where each fixpoint can be computed by 
successive approximation.
The symbolic semi-algorithm {\sf ModelCheck} of Figure~\ref{modelcheck} 
applies this method to infinite-state systems. 

\mypar
Suppose that the input given to {\sf ModelCheck} is the region algebra of a 
symbolic transition system $\sts=(\states,\trans,\regs,\ext{\cdot},\obs)$, a
$\mu$-calculus formula~$\varphi$, and any mapping $\env\:\vars\ra\two^\regs$ 
from the variables to sets of regions.
Then for each recursive call of {\sf ModelCheck}, each~$T_i$, for $i\ge 0$,
is a finite set of regions from~$\regs$, and each recursive call returns a 
finite set of regions from~$\regs$.
It is easy to check that all of these regions are also generated by the 
semi-algorithm {\sf Closure1} on input~$\stt_\sts$.
Thus, if {\sf Closure1} terminates, then so does {\sf ModelCheck}.
Furthermore, if it terminates, then {\sf ModelCheck} returns a set 
$\xsem{\varphi}_\env\subseteq\regs$ of regions such that 
$\bigcup\set{\ext{\reg}\mid\reg\in\xsem{\varphi}_\env}=
  \sem{\varphi}_{\sts,\xenv}$,
where $\xenv(\var)=\bigcup\set{\ext{\reg}\mid\reg\in\env(\var)}$ for all 
$\var\in\vars$.
In particular, if $\varphi$ is closed, then a state $\state\in\states$ 
satisfies $\varphi$ iff $\Member(\state,\reg)$ for some region 
$\reg\in\xsem{\varphi}_\env$.

\begin{figure}[t]
\boxpar{
\begin{verse}
  {\bf Symbolic semi-algorithm} {\sf ModelCheck}\\
  Input: a region algebra $\stt=(\obs,\Pre,\And,\Diff,\Empty)$, 
    a formula $\varphi\in\mlog_1$, and
    a mapping $\env$ with domain~$\vars$.\\[5pt]
  Output: $\xsem{\varphi}_\env$ :=\\
  \qq \IF\ $\varphi=\ob$ \THEN\ \RETURN\ $\set{\ob}$;\\
  \qq \IF\ $\varphi=\dual{\ob}$ \THEN\ \RETURN\ 
    $\set{\Diff(\aob,\ob)\mid\aob\in\obs}$;\\
  \qq \IF\ $\varphi=(\varphi_1\vee\varphi_2)$ \THEN\ \RETURN\
    $\xsem{\varphi_1}_\env\cup\xsem{\varphi_2}_\env$;\\
  \qq \IF\ $\varphi=(\varphi_1\wedge\varphi_2)$ \THEN\\
  \qqq \RETURN\
    $\set{\And(\reg,\areg)\mid\reg\in\xsem{\varphi_1}_\env\mbox{\ and\ }
    \areg\in\xsem{\varphi_2}_\env}$;\\
  \qq \IF\ $\varphi=\EN\varphi'$ \THEN\ \RETURN\
    $\set{\Pre(\reg)\mid\reg\in\xsem{\varphi'}_\env}$;\\
  \qq \IF\ $\varphi=\AN\varphi'$ \THEN\ \RETURN\
    $\obs\bbs\set{\Pre(\reg)\mid\reg\in(\obs\bbs\xsem{\varphi'}_\env)}$;\\
  \qq \IF\ $\varphi=(\mu \var\qmid\varphi')$ \THEN\\
  \qqq $T_0$ := $\emptyset$;\\
  \qqq \FOR\ $i=0,1,2,\ldots$ \DO\\
  \qqqq $T_{i+1}$ := $\xsem{\varphi'}_{\env[\var\mapsto T_i]}$\\
  \qqqq \UNTIL\ 
    $\bigcup\set{\ext{\reg}\mid\reg\in T_{i+1}}\subseteq
      \bigcup\set{\ext{\reg}\mid\reg\in T_i}$;\\
  \qqq \RETURN\ $T_i$;\\
  \qq \IF\ $\varphi=(\nu \var\qmid\varphi')$ \THEN\\
  \qqq $T_0$ := $\obs$;\\
  \qqq \FOR\ $i=0,1,2,\ldots$ \DO\\
  \qqqq $T_{i+1}$ := $\xsem{\varphi'}_{\env[\var\mapsto T_i]}$\\
  \qqqq \UNTIL\ 
    $\bigcup\set{\ext{\reg}\mid\reg\in T_{i+1}}\supseteq
      \bigcup\set{\ext{\reg}\mid\reg\in T_i}$;\\
  \qqq \RETURN\ $T_i$.
\end{verse}
The {\em pairwise-difference\/} operation $T\bbs T'$ between two finite sets  
$T$ and $T'$ of regions is computed inductively as follows:
\begin{verse}
  $T\bbs\emptyset\eeq T$;\\
  $T\bbs(\set{\areg}\cup T')\eeq\set{\Diff(\reg,\areg)\mid\reg\in T}\bbs T'$.
\end{verse}
The termination test 
$\bigcup\set{\ext{\reg}\mid\reg\in T}\subseteq
  \bigcup\set{\ext{\reg}\mid\reg\in T'}$
is decided by checking that $\Empty(\reg)$ for each region 
$\reg\in(T\bbs T')$. 
}
\caption{Model checking}
\label{modelcheck}
\end{figure}

\mypar
{\bf Theorem 1B.} 
{\it 
For all symbolic transition systems $\sts$ in {\sf STS1} and every 
$\mlog_1$-formula~$\varphi$, the symbolic semi-algorithm {\sf ModelCheck} 
terminates on the region algebra $\stt_\sts$ and the input formula~$\varphi$.
}

\mypar
{\bf Corollary 1B} 
{\it 
The $\mlog_1$ model-checking problem is decidable for the class {\sf STS1} of 
symbolic transition systems.
}

\mypar
{\bf Remark: Duality} 
Model checking of $\mlog_1$-formulas on {\sf STS1} systems can also be 
performed by the dual of the semi-algorithm {\sf ModelCheck}.
Suppose that the input given to the dual semi-algorithm 
$\dual{\sf ModelCheck}$ is the dual region algebra of a symbolic 
transition system $\sts=(\states,\trans,\regs,\ext{\cdot},\obs)$, and the
$\mlog_1$-formula~$\varphi$.
If $\sts$ belongs to the class {\sf STS1}, then $\dual{\sf ModelCheck}$
terminates with the output $T\subseteq\regs$ such that 
$\sem{\varphi}_\sts=\bigcap\set{\ext{\reg}\mid\reg\in T}$.
\qed

\begin{figure}[t]
\centering{
\setlength{\unitlength}{0.00083333in}
\begingroup\makeatletter\ifx\SetFigFont\undefined%
\gdef\SetFigFont#1#2#3#4#5{%
  \reset@font\fontsize{#1}{#2pt}%
  \fontfamily{#3}\fontseries{#4}\fontshape{#5}%
  \selectfont}%
\fi\endgroup%
{
\begin{picture}(5183,789)(0,-10)
\put(1433,446){\ellipse{300}{300}}
\put(2625,462){\ellipse{300}{300}}
\put(3825,462){\ellipse{300}{300}}
\put(5025,462){\ellipse{300}{300}}
\path(675,462)(1275,462)
\path(1155.000,432.000)(1275.000,462.000)(1155.000,492.000)
\path(1575,462)(2475,462)
\path(2355.000,432.000)(2475.000,462.000)(2355.000,492.000)
\path(2775,462)(3675,462)
\path(3555.000,432.000)(3675.000,462.000)(3555.000,492.000)
\path(3975,462)(4875,462)
\path(4755.000,432.000)(4875.000,462.000)(4755.000,492.000)
\path(1425,312)(1425,12)(5100,12)(5100,312)
\path(5130.000,192.000)(5100.000,312.000)(5070.000,192.000)
\path(2625,312)(2625,162)(5025,162)(5025,312)
\path(5055.000,192.000)(5025.000,312.000)(4995.000,192.000)
\path(1425,612)(1425,762)(3825,762)(3825,612)
\path(3795.000,732.000)(3825.000,612.000)(3855.000,732.000)
\put(0,462){\makebox(0,0)[lb]{\smash{{{\SetFigFont{12}{14.4}{\rmdefault}{\mddefault}{\updefault}  .  .  .}}}}}
\put(1425,387){\makebox(0,0)[b]{\smash{{{\SetFigFont{12}{14.4}{\rmdefault}{\mddefault}{\updefault}q}}}}}
\put(2625,387){\makebox(0,0)[b]{\smash{{{\SetFigFont{12}{14.4}{\rmdefault}{\mddefault}{\updefault}q}}}}}
\put(3825,387){\makebox(0,0)[b]{\smash{{{\SetFigFont{12}{14.4}{\rmdefault}{\mddefault}{\updefault}q}}}}}
\put(5025,387){\makebox(0,0)[b]{\smash{{{\SetFigFont{12}{14.4}{\rmdefault}{\mddefault}{\updefault}p}}}}}
\end{picture}
}
}
\caption{A symbolic transition system for which {\sf ModelCheck} terminates for every $\varphi$,
while closure under $\Pre$ (and hence {\sf Closure1}) does not.}
\label{counterexample1}
\end{figure}
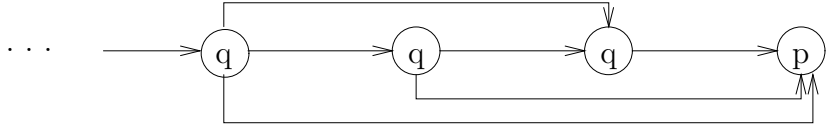

\mypar
{\bf Counterexample}
The converse of Theorem~1B does not hold: there exist symbolic transition
systems $\sts$ such that for every $\mlog_1$-formula $\varphi$,
the symbolic semi-algorithm {\sf Model Check} terminates
on the region algebra $\stt_\sts$ and $\varphi$, and yet $\sts$ is not
in {\sf STS1}.
Indeed, the example of Figure~\ref{counterexample1} shows a symbolic
transition system for which {\sf ModelCheck} terminates for every
formula $\varphi$ of $\mlog_1$, but iteration of $\Pre$ does
not terminate.
In fact, this is true for every transition system whose transition 
relation is transitive.
\qed

\subsection{Example: Singular Hybrid Automata}

The fundamental theorem of timed automata~\cite{AlurDill94} shows that for 
every timed automaton, the (time-abstract) bisimilarity relation has finite 
index.
The proof can be extended to the singular automata~\cite{DES94j}.
It follows that the symbolic semi-algorithm {\sf ModelCheck}, which has been 
implemented for polyhedral hybrid automata in the tool \hytech~\cite{RTSS95}, 
decides all $\mlog_1$ model-checking questions for singular automata.
The singular automata form a maximal class of hybrid automata in
{\sf STS1}.
This is because there is a 2D (two-dimensional) rectangular automaton whose 
bisimilarity relation is state equality~\cite{ICALP95H}.

\mypar
{\bf Theorem 1C} 
{\it 
The singular automata belong to the class {\sf STS1}.
There is a 2D rectangular automaton that does not belong to {\sf STS1}.
}

\def\pc{{\it pc}}
\def\variable{\mathbf{var}}
\subsection{Example: The 2-Process Bakery Protocol}

\begin{figure}[t]
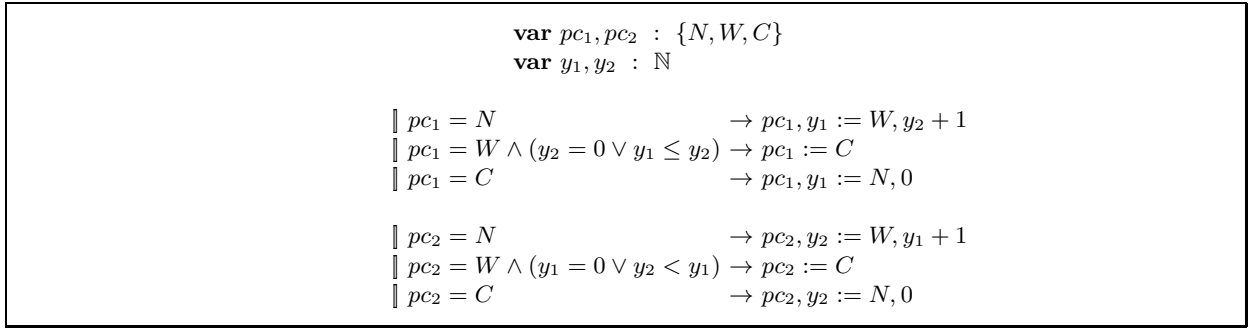

\boxpar{
\begin{verse}
\begin{cmtab}{l}
 \variable\  pc_1, pc_2\ :\ \set{N, W, C}\\
 \variable\  y_1, y_2\ :\ \nats\\
\end{cmtab}

\begin{cmtab}{l}
\begin{chtab}
pc_1=N & pc_1, y_1 := W, y_2+1\\
pc_1 = W \wedge (y_2=0\vee y_1\leq y_2) & pc_1:= C\\
pc_1 = C & pc_1, y_1 := N, 0\\
\end{chtab}
\end{cmtab}

\begin{cmtab}{l}
\begin{chtab}
pc_2=N & pc_2, y_2 := W, y_1+1\\
pc_2 = W \wedge (y_1=0\vee y_2< y_1) & pc_2:= C\\
pc_2 = C & pc_2, y_2 := N, 0\\
\end{chtab}
\end{cmtab}

\end{verse}
}

\caption{The 2-process bakery mutual exclusion algorithm}
\label{bakery}
\end{figure}

Consider the 2-process bakery protocol \cite{Lamport74} for mutual exclusion
presented as a finite collection of guarded commands in Figure~\ref{bakery}.
As presented, the protocol uses two variables (the ``tokens'') that range
over the natural numbers.
The state of the protocol is given by a 4-tuple $(\pc_1, \pc_2, y_1, y_2)$
denoting the values of the program counters in the two processes, and the
values of the tokens $y_1$ and $y_2$.
The observables are boolean formulae over the values of the program counter.
However, we can show that the bisimilarity relation of this transition system
has finite index.
Indeed, define the relation $\cong$ between states of the protocol
as $\state\cong \astate$ iff (1)~$\state(\pc_i) = \astate(\pc_i)$ for $i=1,2$ (where
$\state(\var)$ denotes the valuation to variable $\var$ in state $\state$);
(2)~$\state(y_i)=0$ iff $\astate(y_i)=0$ for $i=1,2$; and
(3)~$\state(y_1)\leq \state(y_2)$ iff $\astate(y_1) \leq \astate(y_2)$.
By a simple case enumeration, it can be seen that $\cong$ is a bisimulation
relation on the state space.
Moreover, the relation has a finite index (the number of equivalence classes is
$72$).
Thus, the 2-process bakery protocol is in {\sf STS1}.
By Theorem~1A, the closure algorithm {\sf Closure1} will terminate on the region
algebra of the 2-process bakery mutual exclusion protocol.

\section{Class-2 Symbolic Transition Systems}

Class-2 systems are characterized by finite similarity quotients.
The region algebra of a class-2 system has a finite subalgebra that contains
the observables and is closed under $\Pre$ and $\And$ operations.
This enables the model checking of all existential and universal 
$\mu$-calculus properties.
Infinite-state examples of class-2 systems are provided by the 2D rectangular
hybrid automata.

\subsection{Finite Characterization: Similarity}

{\bf Definition: Similarity}
Let $\sts$ be a transition system.
Two states $\state$ and $\astate$ of $\sts$ are {\em similar}, denoted
$\state\cong_2^\sts\astate$, if there are simulations $\preceq_1$, $\preceq_2$ on $\sts$ 
such that $\state\preceq_1\astate$ and $\astate\preceq_2\state$.
The state equivalence $\cong_2$ is called {\em similarity}.
\qed

\mypar
{\bf Definition: Class {\sf STS2}}
A symbolic transition system $\sts$ belongs to the class {\sf STS2} if the 
similarity relation $\cong_2^\sts$ has finite index.
\qed

\mypar
Since similarity is coarser than bisimilarity~\cite{vanGlabbeek90}, the class 
{\sf STS2} of symbolic transition systems is a proper extension of 
{\sf STS1}.

\subsection{Symbolic State-space Exploration: Intersection Refinement}

The symbolic semi-algorithm {\sf Closure2} of Figure~\ref{closure2} is an 
abstract version of the method presented in \cite{FOCS95} for computing the 
similarity relation of an infinite-state system.
Suppose that the input given to {\sf Closure2} is the region algebra of a 
symbolic transition system $\sts=(\states,\trans,\regs,\ext{\cdot},\obs)$.
Given two states $\state,\astate\in\states$, we say that $\astate$
{\em simulates\/} $\state$ if $\state\preceq\astate$ for some simulation 
$\preceq$ on~$\sts$.
For $i\ge 0$ and $\state\in\states$, define 
  $$\Sim_i(\state)\eeq
    \bigcap\set{\ext{\reg}\mid\reg\in T_i\mbox{ and }\state\in\ext{\reg}},$$
where the set $T_i$ of regions is computed by {\sf Closure2}.
By induction it is easy to check that for all $i\ge 0$, if $\astate$ 
simulates~$\state$, then $\astate\in\Sim_i(\state)$.
Thus, the extension of every region in $T_i$ is a $\cong_2^\sts$ block, and 
if $\cong_2^\sts$ has finite index, then {\sf Closure2} terminates.
Conversely, suppose that {\sf Closure2} terminates with 
$\ext{T_{i+1}}\subseteq\ext{T_i}$.
{F}rom the definition of simulations it follows that if 
$\astate\in\Sim_i(\state)$, then $\astate$ simulates~$\state$.
This implies that $\cong_2^\sts$ has finite index.

\begin{figure}[t]
\boxpar{
\begin{verse}
  {\bf Symbolic semi-algorithm} {\sf Closure2}\\
  Input: a region algebra $\stt=(\obs,\Pre,\And,\Diff,\Empty)$.\\[5pt]
  \q $T_0$ := $\obs$;\\
  \q \FOR\ $i=0,1,2,\ldots$ \DO\\
  \qq $T_{i+1}$ := $T_i$\\
  \qqq $\cup\ \set{\Pre(\reg)\mid\reg\in T_i}$\\
  \qqq $\cup\ \set{\And(\reg,\areg)\mid\reg,\areg\in T_i}$\\
  \qq \UNTIL\ $\ext{T_{i+1}}\subseteq\ext{T_i}$.
\end{verse}
The termination test $\ext{T_{i+1}}\subseteq\ext{T_i}$ is decided as in 
Figure~\ref{closure1}.
}
\caption{Intersection refinement}
\label{closure2}
\end{figure}

\mypar
{\bf Theorem 2A}
{\it
For all symbolic transition systems~$\sts$, the symbolic semi-algorithm 
{\sf Closure2} terminates on the region algebra $\stt_\sts$ iff $\sts$ 
belongs to the class {\sf STS2}.
}

\mypar
{\bf Corollary 2A} 
{\it 
The $\cong_2$ (similarity) equivalence problem is decidable for the class 
{\sf STS2} of symbolic transition systems.
}

\subsection{Decidable Properties: Negation-free Branching Time}

{\bf Definition: Negation-free $\mu$-calculus}
The {\em negation-free $\mu$-calculus\/} consists of the $\mu$-calculus
formulas that are generated by the grammar
  \[\varphi\ ::=\
    \ob \mid
    \var \mid 
    \varphi\vee\varphi \mid 
    \varphi\wedge\varphi \mid 
    \EN\varphi \mid 
    (\mu\var\qmid\varphi) \mid 
    (\nu\var\qmid\varphi),
    \]
for constants $\ob\in\consts$ and variables $\var\in\vars$.
The state logic $\mlog_2$ consists of the closed formulas of the negation-free
$\mu$-calculus.
The state logic $\dual{\mlog_2}$ consists of the duals of all
$\mlog_2$-formulas. 
\qed 

\mypar
The following facts about the negation-free $\mu$-calculus and its dual are 
relevant in our context~\cite{CavBook}.
First, both $\mlog_2$ and $\dual{\mlog_2}$ admit abstraction, and the state 
equivalence induced by both $\mlog_2$ and $\dual{\mlog_2}$ is~$\cong_2$ 
(similarity).
It follows that the logic $\mlog_1$ with negation is more expressive than 
either $\mlog_2$ or~$\dual{\mlog_2}$.
Second, the negation-free logic $\mlog_2$ is more expressive than the 
existential fragments of $\ctlstar$ and $\ctl$, which also induce 
similarity, and the dual logic $\dual{\mlog_2}$ is more expressive than the 
universal fragments of $\ctlstar$ and $\ctl$, which again induce similarity. 

\mypar
If we apply the symbolic semi-algorithm {\sf ModelCheck} of 
Figure~\ref{modelcheck} to the region algebra of a symbolic transition system
$\sts$ and an input formula from~$\mlog_2$, then the cases 
$\varphi=\dual{\ob}$ and $\varphi=\AN\varphi'$ are never executed.
It follows that all regions which are generated by {\sf ModelCheck} are also 
generated by the semi-algorithm {\sf Closure2} on input~$\stt_\sts$.
Thus, if {\sf Closure2} terminates, then so does {\sf ModelCheck}.

\mypar
{\bf Theorem 2B} 
{\it 
For all symbolic transition systems $\sts$ in {\sf STS2} and every 
$\mlog_2$-formula~$\varphi$, the symbolic semi-algorithm {\sf ModelCheck} 
terminates on the region algebra $\stt_\sts$ and the input formula~$\varphi$.
}

\mypar
{\bf Corollary 2B} 
{\it 
The $\mlog_2$ and $\dual{\mlog_2}$ model-checking problems are decidable 
for the class {\sf STS2} of symbolic transition systems.
}

\subsection{Example: 2D Rectangular Hybrid Automata}

For every 2D rectangular automaton, the (time-abstract) similarity relation 
has finite index~\cite{FOCS95}.
It follows that the symbolic semi-algorithm {\sf ModelCheck}, as implemented 
in \hytech, decides all $\mlog_2$ and $\dual{\mlog_2}$ model-checking 
questions for 2D rectangular automata.
The 2D rectangular automata form a maximal class of hybrid automata in 
{\sf STS2}.
This is because there is a 3D rectangular automaton whose similarity relation
is state equality~\cite{CONCUR96HK}.

\mypar
{\bf Theorem 2C} 
{\it 
The 2D rectangular automata belong to the class {\sf STS2}.
There is a 3D rectangular automaton that does not belong to {\sf STS2}.
}

\section{Class-3 Symbolic Transition Systems}

Class-3 systems are characterized by finite trace-equivalence quotients.
The region algebra of a class-3 system has a finite subalgebra that contains
the observables and is closed under $\Pre$ operations and those $\And$ 
operations for which one of the two arguments is an observable. 
This enables the model checking of all linear temporal properties.
Infinite-state examples of class-3 systems are provided by the rectangular
hybrid automata.

\subsection{Finite Characterization: Traces}

{\bf Definition: Trace equivalence}
Let $\sts=(\states,\trans,\cdot,\ext{\cdot},\obs)$ be a transition system.
Given a state $\state_0\in\states$, a {\em source-$\state_0$ trace\/}
$\trace$ of $\sts$ is a finite or infinite sequence $\ob_0\ob_1\ldots$ of 
observables $\ob_i\in\obs$ such that 
\begin{verse}
  1. $\state_0\in\ext{\ob_0}$;\\
  2. for all $0\le i$, there is a state 
    $\state_{i+1}\in(\trans(\state_i)\cap\ext{\ob_{i+1}})$.
\end{verse}
If the trace is a finite sequence $\ob_0\ob_1\ldots\ob_n$, 
the number $n$ of observables (minus~1) is called the {\em length\/} of the 
trace~$\trace$, the final state $\state_n$ is the {\em sink\/} of~$\trace$, 
and the final observable $\ob_n$ is the {\em target\/} of~$\trace$. 
The length of an infinite trace is infinity.
Two states $\state,\astate\in\states$ are {\em trace equivalent}, denoted
$\state\cong_3^\sts\astate$, if every source-$\state$ trace of $\sts$ is a 
source-$\astate$ trace of~$\sts$, and vice versa.
The state equivalence $\cong_3$ is called {\em trace equivalence}.
Two states $\state,\astate\in\states$ are {\em finite trace equivalent}, denoted 
$\state\cong_{3f}^\sts\astate$, if every finite source-$\state$ trace of $\sts$
is a source-$\astate$ trace of~$\sts$, and vice versa.
The state equivalence $\cong_{3f}$ is called {\em finite trace equivalence}.
\qed

\mypar
{\bf Definition: Class {\sf STS3}}
A symbolic transition system $\sts$ belongs to the class {\sf STS3} if the 
trace-equivalence relation $\cong_3^\sts$ has finite index.
\qed

\mypar
Since trace equivalence is coarser than similarity~\cite{vanGlabbeek90}, the 
class {\sf STS3} of symbolic transition systems is a proper extension of 
{\sf STS2}.

\subsection{Symbolic State-space Exploration: Observation Refinement}

Trace equivalence can be characterized operationally by the symbolic
semi-algorithm {\sf Closure3} of Figure~\ref{closure3}.
We shall show that, when the input is the region algebra of a symbolic
transition system $\sts=(\states,\trans,\regs,\ext{\cdot},\obs)$, then
{\sf Closure3} terminates iff the trace-equivalence relation $\cong_3^\sts$
has finite index.
Furthermore, upon termination, $\state\cong_3^\sts\astate$ iff for each 
region $\reg\in T_i$, we have $\state\in\ext{\reg}$ iff 
$\astate\in\ext{\reg}$.

\begin{figure}[t]
\boxpar{
\begin{verse}
  {\bf Symbolic semi-algorithm} {\sf Closure3}\\
  Input: a region algebra $\stt=(\obs,\Pre,\And,\Diff,\Empty)$.\\[5pt]
  \q $T_0$ := $\obs$;\\
  \q \FOR\ $i=0,1,2,\ldots$ \DO\\
  \qq $T_{i+1}$ := $T_i$\\
  \qqq $\cup\ \set{\Pre(\reg)\mid\reg\in T_i}$\\
  \qqq $\cup\ \set{\And(\reg,\ob)\mid\reg\in T_i\mbox{\ and\ }\ob\in\obs}$\\
  \qq \UNTIL\ $\ext{T_{i+1}}\subseteq\ext{T_i}$.
\end{verse}
The termination test $\ext{T_{i+1}}\subseteq\ext{T_i}$ is decided as in 
Figure~\ref{closure1}.
}
\caption{Observation refinement}
\label{closure3}
\end{figure}

\mypar
{\bf Theorem 3A}
{\it
For all symbolic transition systems~$\sts$, the symbolic semi-algorithm 
{\sf Closure3} terminates on the region algebra $\stt_\sts$ iff $\sts$ 
belongs to the class {\sf STS3}.
}

\mypar
{\bf Proof}
We proceed in two steps.
First, we show that {\sf Closure3} terminates on the region algebra 
$\stt_\sts$ iff the equivalence relation $\smash{\cong^\sts_{\mlog_3}}$ 
induced by the deterministic $\mu$-calculus (defined below) has finite index.
Second, we show that $\smash{\cong_{\mlog_3}}$ coincides with trace 
equivalence.
The proof of the first part proceeds as usual.
It can be seen by induction that for all $i\ge 0$, the extension of every
region in~$T_i$, as computed by {\sf Closure3}, is a 
$\smash{\cong^\sts_{\mlog_3}}$ block.
Thus, if $\smash{\cong^\sts_{\mlog_3}}$ has finite index, then {\sf Closure3}
terminates.
Conversely, suppose that {\sf Closure3} terminates with 
$\ext{T_{i+1}}\subseteq\ext{T_i}$.
It can be shown that if two states are not 
$\smash{\cong^\sts_{\mlog_3}}$-equivalent, then there is a region in $T_i$ 
which contains one state but not the other. 
It follows that if for each region $\reg\in T_i$, we have 
$\state\in\ext{\reg}$ iff $\astate\in\ext{\reg}$, then
$\state\smash{\cong^\sts_{\mlog_3}}\astate$.
This implies that $\smash{\cong^\sts_{\mlog_3}}$ has finite index.

\mypar
For the second part, we show that $\mlog_3$ is as expressive as the logic 
$\EB$, whose formulas are the existentially interpreted B\"uchi automata,
and $\EB$ is as expressive as $\mlog_3$.
This result is implicit in a proof by~\cite{EmersonJutlaSistla93}.
We recall a few definitions.
A \textit{B\"uchi automaton} $\Buchi$ is a tuple 
$(S, \Phi, \rightarrow, s_0, F)$, where $S$ is a finite set of states, 
$\Phi$ is a finite input alphabet, 
$\rightarrow \;\subseteq S\times \Phi \times S$ is the transition relation,
$s_0\in S$ is the start state, 
and $F\subseteq S$ is the set of B\"uchi accepting states.
An \textit{execution} of $\Buchi$ on an $\omega$-word 
$w = w_0w_1\ldots\in\Phi^\omega$ 
is an infinite sequence $r = s_0s_1\ldots$ of states in $S$, starting from the 
initial state~$s_0$, such that
$s_i\smash{\stackrel{w_i}{\rightarrow}} s_{i+1}$ for all $i\ge 0$.
The execution $r$ is \textit{accepting} if some state in $F$ occurs infinitely 
often in~$r$.
The automaton $\Buchi$ \textit{accepts} the word $w$ if it has an accepting
execution on~$w$.
The \textit{language} $L(\Buchi)\subseteq\Phi^\omega$ is the set of 
$\omega$-words accepted by~$\Buchi$.

The proof is based on the following constructions.
By induction on the structure of an $\mlog_3$-formula~$\varphi$, we can 
construct a B\"uchi automaton $B_\varphi$ such that for all transition 
systems~$\sts$, a state $\state$ of $\sts$ satisfies $\varphi$ iff for some 
infinite source-$\state$ trace of $\sts$ is accepted by~$B_\varphi$.
Conversely, given a B\"uchi automaton~$B$, we construct an 
$\mlog_3$-formula which is equivalent to $\exists B$.
Let $\Buchi$ be a B\"uchi automaton.
For notational convenience, we present the formula 
in equational form \cite{CKS92};
it can be easily converted to the standard
representation by unrolling the equations, and binding variables with $\mu$
or $\nu$-fixpoints.
For each set $R\in 2^\obs$, let $\psi_R$ abbreviate the formula
$\bigwedge R\wedge \bigwedge \set{\dual{\ob}\mid\ob\in\obs\backslash R}$.
For each state $s$ of $\Buchi$, we introduce a propositional variable $X_s$.
The equation for $X_s$ is 
\[
  X_s\ =_\lambda\ \bigvee \{\psi_R\wedge \exists\!\Next X_{s'}\mid 
				s \stackrel{R}{\rightarrow} s'\},
\]
where $\lambda = \nu$ if $s\in F$ is an accepting state, 
and $\lambda=\mu$ otherwise.
The top-level variable is $X_{s_0}$, where $s_0$ is the initial state.
The correctness of the procedure follows from \cite{BC96}.
An equivalent construction is given in \cite{Dam94}.

Since the state equivalence induced by $\EB$ is trace equivalence, it follows
that $\smash{\cong_{\mlog_3}}$ is also trace equivalence.
\qed

\mypar
{\bf Corollary 3A} 
{\it 
The $\cong_3$ (trace) equivalence problem is decidable for the class 
{\sf STS3} of symbolic transition systems.
}

\subsection{Decidable Properties: Linear Time}

{\bf Definition: Deterministic $\mu$-calculus}
The {\em deterministic $\mu$-calculus\/} 
(also called ``$L_1$'' in~\cite{EmersonJutlaSistla93})
consists of the $\mu$-calculus formulas that are generated by the grammar
  \[\varphi\ ::=\
    \ob \mid
    \var \mid 
    \varphi\vee\varphi \mid 
    \ob\wedge\varphi \mid 
    \EN\varphi \mid 
    (\mu\var\qmid\varphi) \mid 
    (\nu\var\qmid\varphi),
    \]
for constants $\ob\in\consts$ and variables $\var\in\vars$.
The state logic $\mlog_3$ consists of the closed formulas of the deterministic
$\mu$-calculus.
The state logic $\dual{\mlog_3}$ consists of the duals of all 
$\mlog_3$-formulas. 
\qed 

\mypar
The following facts about the deterministic $\mu$-calculus and its dual are 
relevant in our context (cf.~the second part of the proof of Theorem~3A).
First, both $\mlog_3$ and $\dual{\mlog_3}$ admit abstraction, and the state 
equivalence induced by both $\mlog_3$ and $\dual{\mlog_3}$ is~$\cong_3$ 
(trace equivalence).
It follows that the logic $\mlog_2$ with unrestricted conjunction is more 
expressive than~$\mlog_3$, and $\dual{\mlog_2}$ is more expressive 
than~$\dual{\mlog_3}$.
Second, the logic $\mlog_3$ with restricted conjunction is more expressive 
than the existential interpretation of the linear temporal logic $\ltl$, 
which also induces trace equivalence.
For example, the existential $\ltl$ formula $\exists(\ob\U\aob)$ 
(``on some trace, $\ob$ until~$\aob$'') is equivalent to the 
$\mlog_3$-formula $(\mu\var\qmid\aob\lor(\ob\wedge\EN\var))$
(notice that one argument of the conjunction is a constant).
The dual logic $\dual{\mlog_3}$ is more expressive than the usual, 
universal interpretation of $\ltl$, which again induces trace equivalence.
For example, the (universal) $\ltl$ formula $\ob\W\aob$ 
(``on all traces, either $\ob$ forever, or $\ob$ until~$\aob$'') is 
equivalent to the $\dual{\mlog_3}$-formula
$(\nu\var\qmid\ob\land\AN(\aob\vee\var))$
(notice that one argument of the disjunction is a constant).

\mypar
If we apply the symbolic semi-algorithm {\sf ModelCheck} of 
Figure~\ref{modelcheck} to the region algebra of a symbolic transition system
$\sts$ and an input formula from~$\mlog_3$, then all regions which are 
generated by {\sf ModelCheck} are also generated by the semi-algorithm 
{\sf Closure3} on input~$\stt_\sts$.
Thus, if {\sf Closure3} terminates, then so does {\sf ModelCheck}.

\mypar
{\bf Theorem 3B}
{\it
For all symbolic transition systems $\sts$ in {\sf STS3} and every 
$\mlog_3$-formula~$\varphi$, the symbolic semi-algorithm {\sf ModelCheck} 
terminates on the region algebra $\stt_\sts$ and the input formula~$\varphi$.
}

\mypar
{\bf Corollary 3B} 
{\it 
The $\mlog_3$ and $\dual{\mlog_3}$ model-checking problems are decidable 
for the class {\sf STS3} of symbolic transition systems.
}

\mypar
{\bf Remark: $\ltl$ model checking}
These results suggest, in particular, a symbolic procedure for model 
checking $\ltl$ properties over {\sf STS3} systems~\cite{TACAS00}.
Suppose that $\sts$ is a symbolic transition system in the class {\sf STS3}, 
and $\varphi$ is an $\ltl$ formula.
First, convert $\neg\varphi$ to a B\"uchi automaton $\Buchi_{\neg\varphi}$ using a 
tableau construction, and then to an equivalent $\mlog_3$-formula~$\psi$
(introduce one variable per state of~$\Buchi_{\neg\varphi}$).
Second, run the symbolic semi-algorithm {\sf ModelCheck} on inputs
$\stt_\sts$ and~$\psi$.
It will terminate with a representation of the complement of the set of
states that satisfy $\varphi$ in~$\sts$.

While {\sf ModelCheck} provides a symbolic semi-algorithm for $\ltl$,
traditionally, a different method is used for symbolic model checking of 
\ltl\/ formulas \cite{CGL94}.
Given a state $\state$ of a finite-state transition structure $\sts$, and an \ltl\/
formula $\varphi$, the model-checking question for \ltl\/ can be solved by
constructing the product of $\sts$ with the tableau automaton
$\Buchi_\varphi$, and then checking the nonemptiness of a B\"uchi condition
on the product structure.
A B\"uchi condition is an \ltl\/ formula of the form $\Box\Diamond\psi$,
where $\psi$ is a disjunction of observables;
therefore nonemptiness can be checked symbolically by evaluating the 
equivalent formula 
  $$\chi\ =\ 
  \nu X_1.\;\mu X_2.\;(\EN X_2\vee(\psi\wedge\EN X_1))$$
of $\mlog_3$.

To extend this method to infinite-state structures, we need to be more
formal. 
Let $\sts = (\states,\trans,\regs,\ext{\cdot},\obs)$ be a symbolic transition system and let
$\Buchi_\varphi = (S, 2^{\Pi}, \rightarrow, s_0, F)$ be a tableau automaton.
The {\em product structure\/} 
$\sts_\varphi=(S\times \states,\trans_\varphi,S\times \regs,\ext{\cdot}_\varphi,\obs_\varphi)$ 
is defined as follows.
The set of states of $\sts_\varphi$ is the Cartesian product $S\times \states$, and
the set of regions of $\sts_\varphi$ is the Cartesian product
$S\times \regs$. 
The extension $\ext{(s\reg)}_\varphi$
for the region $(s, \reg)$ is
the set of states $\set{s}\times \ext{\reg}$.
The set of observables $\obs_\varphi$ is  $S \times \obs$,
for an observable $(s,\ob)\in\obs_\varphi$, 
define $(s',\state)\in \ext{(s,\ob)}_\varphi$ iff $s'=s$ and $\state\in\ext{\ob}$;
that is, the state of the tableau automaton is also observable.
Define $(s',\astate)\in\trans_\varphi(s,\state)$ iff 
$s\stackrel{\ob}{\rightarrow}s'$ and $\astate\in\trans(\state)$ and $\state\in\ext{\ob}$.
Then $\state\in\sem{\varphi}_\sts$, for $\state\in \states$, iff
$(s_0,\state)\in\sem{\Box\Diamond\psi}_{\sts_\varphi}$, 
where $\psi=\bigvee_{s\in F,\ob\in\obs}(s,\ob)$.
Since the tableau automaton $\Buchi_\varphi$ is finite, it is easy to check 
that $\stt_\varphi$, with the extension function $\ext{\cdot}_\varphi$, is a region algebra for
$\sts_\varphi$. 
Let {\sf AutomLTL} be the \textit{product-automaton based algorithm for \ltl\/ model
checking\/} which,
given an \ltl\/ formula $\varphi$ and a symbolic transition system $\sts$, 
evaluates the $\mlog_3$ formula $\chi$ (representing a B\"uchi condition) 
on the product system $\sts_{\lnot\varphi}$ (using the semi-algorithm {\sf ModelCheck}).
It is not difficult to see that if observation refinement terminates on
$\sts$ in $k$ steps, then it also terminates on $\sts_\varphi$ in 
$k$ steps 
(if {\sf Closure3} generates $m$ regions on $\sts$, then it generates at most 
$m\cdot |S|$ regions on $\sts_\varphi$).

\mypar
{\bf Corollary 3B$'$}\label{coro-automatonbasedltl}
{\it
For all symbolic transition systems $\sts$ in {\sf STS3}, and every \ltl\/ 
formula $\varphi$, the symbolic semi-algorithm {\sf AutomLTL} 
terminates on the region algebra $\stt_\sts$ and the input formula~$\varphi$.
}

\mypar
Indeed, by induction on the construction of regions, one can show that
for each region representative $(s,\reg)$ computed in the product-automaton 
based algorithm, the variable $X_s$ in the $\mu$-calculus based algorithm 
represents the region $\ext{\reg}$ at some stage of the computation, and 
conversely, for each valuation $R$ of the variable $X_s$ in the 
$\mu$-calculus based algorithm, a region representative of $\set{s}\times R$ 
is computed in the product-automaton based algorithm.
Thus, the two methods are equivalent in the regions they generate.
\qed
 
\mypar
{\bf Remark: Finite Trace Equivalence}
Let {\sf STS3f} be the class of symbolic transition systems whose
finite trace equivalence relation has finite index.

\mypar
{\bf Definition: Finitary Deterministic $\mu$-calculus}
The {\em finitary fragment} of the deterministic $\mu$-calculus
consists of the formulas of the deterministic $\mu$-calculus without
the greatest fixpoint operator.
Formally, formulas are generated by the grammar
  \[\varphi\ ::=\
    \ob \mid
    \var \mid 
    \varphi\vee\varphi \mid 
    \ob\wedge\varphi \mid 
    \EN\varphi \mid 
    (\mu\var\qmid\varphi),
    \]
for constants $\ob\in\consts$ and variables $\var\in\vars$.
The state logic $\mlog_{3f}$ consists of the closed formulas of the finitary deterministic
$\mu$-calculus.
The state logic $\dual{\mlog_{3f}}$ consists of the duals of all 
$\mlog_{3f}$-formulas. 
\qed

\mypar
From the proof of Theorem~3A, we notice that the finitary deterministic $\mu$-calculus
is equally expressive as the logic $\EA$ whose formulas are the existentially
interpreted finite automata, in other words, $\mlog_{3f}$ expresses exactly
the regular sets.
Thus the following corollary is immediate.

\mypar
{\bf Corollary 3BFinite}
{\it
For all symbolic transition systems $\sts$ in {\sf STS3f} and every 
$\mlog_{3f}$-formula~$\varphi$, the symbolic semi-algorithm {\sf ModelCheck} 
terminates on the region algebra $\stt_\sts$ and the input formula~$\varphi$.
Hence, the $\mlog_{3f}$ and $\dual{\mlog_3}$ model-checking problems are
decidable for the class {\sf STS3f} of symbolic transition systems.
}

\subsection{Example: Rectangular Hybrid Automata}

For every rectangular automaton, the (time-abstract) trace-equivalence 
relation has finite index~\cite{STOC95j}.
It follows that the symbolic semi-algorithm {\sf ModelCheck}, as implemented 
in \hytech, decides all $\mlog_3$ and $\dual{\mlog_3}$ model-checking 
questions for rectangular automata.
The rectangular automata form a maximal class of hybrid automata in
{\sf STS3}.
This is because for simple generalizations of rectangular automata, the 
reachability problem is undecidable~\cite{STOC95j}.

\mypar
{\bf Theorem 3C} 
{\it 
The rectangular automata belong to the class {\sf STS3}.
}

\section{Class-4 Symbolic Transition Systems}

We define two states of a transition system to be ``distance equivalent'' if
for every distance~$d$, the same observables can be reached in $d$
transitions. 
Class-4 systems are characterized by finite distance-equivalence quotients.
The region algebra of a class-4 system has a finite subalgebra that
contains the observables and is closed under $\Pre$ operations.
This enables the model checking of all existential conjunction-free and 
universal disjunction-free $\mu$-calculus properties, such as the property 
that an observable can be reached in an even number of transitions.

\subsection{Finite Characterization: Equi-distant Targets}

{\bf Definition: Distance equivalence}
Let $\sts$ be a transition system.
Two states $\state$ and $\astate$ of $\sts$ are {\em distance equivalent},
denoted $\state\cong_4^\sts\astate$, if for every source-$\state$ trace of 
$\sts$ with length $n$ and target~$\ob$, there is a source-$\astate$ trace 
of $\sts$ with length $n$ and target~$\ob$, and vice versa.
The state equivalence $\cong_4$ is called {\em distance equivalence}.
\qed

\mypar
{\bf Definition: Class {\sf STS4}}
A symbolic transition system $\sts$ belongs to the class {\sf STS4} if the 
distance-equivalence relation $\cong_4^\sts$ has finite index.
\qed

\begin{figure}[t]
\centering{
\setlength{\unitlength}{2763sp}%
\begingroup\makeatletter\ifx\SetFigFont\undefined%
\gdef\SetFigFont#1#2#3#4#5{%
  \reset@font\fontsize{#1}{#2pt}%
  \fontfamily{#3}\fontseries{#4}\fontshape{#5}%
  \selectfont}%
\fi\endgroup%
\begin{picture}(3766,1964)(293,-1718)
\thinlines
\put(1351,-811){\circle{300}}
\put(3901,-1561){\circle{300}}
\put(3901,-811){\circle{300}}
\put(3001,-811){\circle{300}}
\put(3451, 89){\circle{300}}
\put(451,-811){\circle{300}}
\put(901, 89){\circle{300}}
\put(3931,-961){\vector( 0,-1){450}}
\put(3331, 14){\vector(-1,-2){330}}
\put(1351,-961){\vector( 0,-1){450}}
\put(751, 14){\vector(-1,-2){330}}
\put(1351,-1561){\circle{300}}
\put(1051, 14){\vector( 1,-2){330}}
\put(601, 89){\makebox(0,0)[lb]{\smash{\SetFigFont{9}{10.8}{\rmdefault}{\mddefault}{\updefault}$s$}}}
\put(3631, 14){\vector( 1,-2){330}}
\put(1291,-1621){\makebox(0,0)[lb]{\smash{\SetFigFont{8}{9.6}{\rmdefault}{\mddefault}{\updefault}$q$}}}
\put(1291,-871){\makebox(0,0)[lb]{\smash{\SetFigFont{8}{9.6}{\rmdefault}{\mddefault}{\updefault}$q$}}}
\put(391,-871){\makebox(0,0)[lb]{\smash{\SetFigFont{8}{9.6}{\rmdefault}{\mddefault}{\updefault}$p$}}}
\put(3841,-871){\makebox(0,0)[lb]{\smash{\SetFigFont{8}{9.6}{\rmdefault}{\mddefault}{\updefault}$p$}}}
\put(3391, 29){\makebox(0,0)[lb]{\smash{\SetFigFont{8}{9.6}{\rmdefault}{\mddefault}{\updefault}$p$}}}
\put(2941,-871){\makebox(0,0)[lb]{\smash{\SetFigFont{8}{9.6}{\rmdefault}{\mddefault}{\updefault}$q$}}}
\put(3841,-1621){\makebox(0,0)[lb]{\smash{\SetFigFont{8}{9.6}{\rmdefault}{\mddefault}{\updefault}$q$}}}
\put(841, 29){\makebox(0,0)[lb]{\smash{\SetFigFont{8}{9.6}{\rmdefault}{\mddefault}{\updefault}$p$}}}
\put(3151, 89){\makebox(0,0)[lb]{\smash{\SetFigFont{9}{10.8}{\rmdefault}{\mddefault}{\updefault}$t$}}}
\end{picture}
}
\caption{Distance equivalence is coarser than trace equivalence}
\label{sts3sts4}
\end{figure}
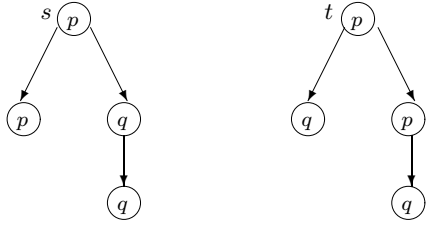

\mypar
Figure~\ref{sts3sts4} shows that distance equivalence is coarser than 
trace equivalence ($\state$~and $\astate$ are distance equivalent but not 
trace equivalent).
It follows that the class {\sf STS4} of symbolic transition systems is a 
proper extension of {\sf STS3}.

\subsection{Symbolic State-space Exploration: Predecessor Iteration}

\begin{figure}[t]
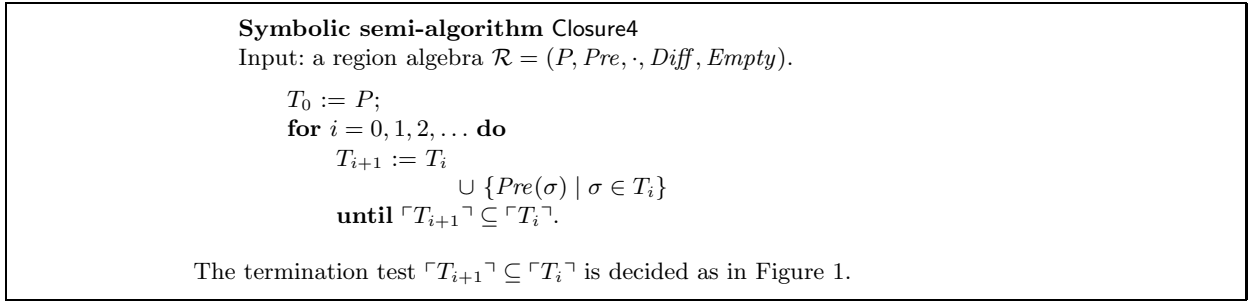

\boxpar{
\begin{verse}
  {\bf Symbolic semi-algorithm} {\sf Closure4}\\
  Input: a region algebra $\stt=(\obs,\Pre,\cdot,\Diff,\Empty)$.\\[5pt]
  \q $T_0$ := $\obs$;\\
  \q \FOR\ $i=0,1,2,\ldots$ \DO\\
  \qq $T_{i+1}$ := $T_i$\\
  \qqq $\cup\ \set{\Pre(\reg)\mid\reg\in T_i}$\\
  \qq \UNTIL\ $\ext{T_{i+1}}\subseteq\ext{T_i}$.
\end{verse}
The termination test $\ext{T_{i+1}}\subseteq\ext{T_i}$ is decided as in 
Figure~\ref{closure1}.
}
\caption{Predecessor iteration}
\label{closure4}
\end{figure}

The symbolic semi-algorithm {\sf Closure4} of Figure~\ref{closure4}
computes the subalgebra of a region algebra $\stt_\sts$ that contains
the observables and is closed under the $\Pre$ operation.
Suppose that the input given to {\sf Closure4} is the region algebra of a 
symbolic transition system $\sts=(\states,\trans,\regs,\ext{\cdot},\obs)$.
For $i\ge 0$ and $\state,\astate\in\states$, define 
$\state\sim_i^\sts\astate$ if for every source-$\state$ trace of $\sts$ with 
length $n\le i$ and target~$\ob$, there is a source-$\astate$ trace of $\sts$
with length $n$ and target~$\ob$, and vice versa.
By induction it is easy to check that for all $i\ge 0$, the extension of 
every region in~$T_i$, as computed by {\sf Closure4}, is a 
$\sim_i^\sts$ block.
Since $\sim_i^\sts$ is as coarse as $\sim_{i+1}^\sts$ for all $i\ge 0$, and 
$\cong_2^\sts$ is equal to $\bigcap\set{\sim^\sts_i\mid i\ge 0}$, if 
$\cong_2^\sts$ has finite index, then $\cong_2^\sts$ is equal to 
$\sim_i^\sts$ for some $i\ge 0$. 
Then, {\sf Closure2} will terminate in $i$ iterations.
Conversely, suppose that {\sf Closure4} terminates with 
$\ext{T_{i+1}}\subseteq\ext{T_i}$. 
In this case, if for all regions $\reg\in T_i$, we have 
$\state\in\ext{\reg}$ iff $\astate\in\ext{\reg}$, then
$\state\cong_4^\sts\astate$.
This is because if $\state$ can reach an observable $\ob$ in $n$ transitions, 
but $\astate$ cannot, then there is a region in~$T_i$, namely, 
$\Pre^n(\ob)$, such that $\state\in\ext{\Pre^n(\ob)}$ and
$\astate\not\in\ext{\Pre^n(\ob)}$.
It follows that $\cong_4^\sts$ has finite index.

\mypar
{\bf Theorem 4A}
{\it
For all symbolic transition systems~$\sts$, the symbolic semi-algorithm 
{\sf Closure4} terminates on the region algebra $\stt_\sts$ iff $\sts$ 
belongs to the class {\sf STS4}.
}

\mypar
{\bf Corollary 4A} 
{\it 
The $\cong_4$ (distance) equivalence problem is decidable for the class 
{\sf STS4} of symbolic transition systems.
}

\subsection{Decidable Properties: Conjunction-free Linear Time}

{\bf Definition: Conjunction-free $\mu$-calculus}
The {\em conjunction-free $\mu$-calculus\/} consists of the $\mu$-calculus
formulas that are generated by the grammar
  \[\varphi\ ::=\
    \ob \mid
    \var \mid 
    \varphi\vee\varphi \mid 
    \EN\varphi \mid 
    (\mu\var\qmid\varphi) 
    \]
for constants $\ob\in\consts$ and variables $\var\in\vars$.
The state logic $\mlog_4$ consists of the closed formulas of the
conjunction-free $\mu$-calculus.
The state logic $\dual{\mlog_4}$ consists of the duals of all 
$\mlog_4$-formulas. 
\qed 

\mypar
{\bf Definition: Conjunction-free temporal logic}
The formulas of the {\em conjunction-free temporal logic\/} $\tlog_4$ are 
generated by the grammar
  \[\varphi\ ::=\
    \ob \mid
    \varphi\vee\varphi \mid 
    \EN\varphi \mid
    \EDB{\le d}\varphi \mid
    \ED\varphi,
    \]
for constants $\ob\in\consts$ and nonnegative integers~$d$.
Let $\sts=(\states,\trans,\cdot,\ext{\cdot},\obs)$ be a transition system 
whose observables include all constants; 
that is, $\consts\subseteq\obs$.
The $\tlog_4$-formula $\varphi$ defines the set 
$\sem{\varphi}_\sts\subseteq\states$ of satisfying states:
\begin{verse}
  $\sem{\ob}_\sts\eeq\ext{\ob}$;\\
  $\sem{\varphi_1\vee\varphi_2}_\sts\eeq
    \sem{\varphi_1}_\sts\cup\sem{\varphi_2}_\sts$;\\
  $\sem{\EN\varphi}_\sts\eeq
    \set{\state\in\states\mid(\exists\astate\in\trans(\state)
    \qmid\astate\in\sem{\varphi}_\sts)}$;\\
  $\sem{\EDB{\le d}\varphi}_\sts\eeq\{\state\in\states\mid$
    there is a source-$\state$ trace of $\sts$ with\\
    \qqqq length at most $d$ and sink in $\sem{\varphi}_\sts\}$;\\
  $\sem{\ED\varphi}_\sts\eeq
    \set{\state\in\states\mid\mbox{there is a source-$\state$ trace of $\sts$ 
    with sink in\ } \sem{\varphi}_\sts}$.
\end{verse}
(The constructor $\ED_{\le d}$ is definable from $\EN$ and~$\vee$;
however, it will be essential in the $\EN$-free fragment of $\tlog_4$ we 
will consider below.)
\qed

\mypar
{\bf Remark: Duality} 
For every $\tlog_4$-formula~$\varphi$, the {\em dual\/} formula 
$\dual{\varphi}$ is obtained by replacing the constructors $\ob$, $\vee$,
$\EN$, $\ED_{\le d}$, and $\ED$ by $\dual{\ob}$, $\wedge$, $\AN$, 
$\AB_{\le d}$, and $\AB$, respectively.
The semantics of the dual constructors is defined as usual, such that
$\sem{\dual{\varphi}}_\sts=\states\bs\sem{\varphi}_\sts$.
The state logic $\dual{\tlog_4}$ consists of the duals of all 
$\tlog_4$-formulas. 
It follows that the answer of the model-checking question for a state 
$\state\in\states$ and an $\dual{\tlog_4}$-formula $\dual{\varphi}$ 
is complementary to the answer of the model-checking question for $\state$ 
and the $\tlog_4$-formula~$\varphi$.
\qed

\mypar
The following facts about the conjunction-free $\mu$-calculus,
conjunction-free temporal logic, and their duals are relevant in our
context. 
First, both $\mlog_4$ and $\dual{\mlog_4}$ admit abstraction, and the state 
equivalence induced by both $\mlog_4$ and $\dual{\mlog_4}$ is~$\cong_4$ 
(distance equivalence).
It follows that the logic $\mlog_3$ with restricted conjunction is more 
expressive than~$\mlog_4$, and $\dual{\mlog_3}$ is more expressive 
than~$\dual{\mlog_4}$.
Second, the conjunction-free $\mu$-calculus $\mlog_4$ is more expressive than 
the conjunction-free temporal logic~$\tlog_4$, and $\dual{\mlog_4}$ is more 
expressive than~$\dual{\tlog_4}$, both of which also induce distance 
equivalence.
For example, the property that an observable can be reached in an even number
of transitions can be expressed in $\mlog_4$ but not in~$\tlog_4$.

\mypar
If we apply the symbolic semi-algorithm {\sf ModelCheck} of 
Figure~\ref{modelcheck} to the region algebra of a symbolic transition system
$\sts$ and an input formula from~$\mlog_4$, then all regions which are 
generated by {\sf ModelCheck} are also generated by the semi-algorithm 
{\sf Closure4} on input~$\stt_\sts$.
Thus, if {\sf Closure4} terminates, then so does {\sf ModelCheck}.

\mypar
{\bf Theorem 4B}
{\it
For all symbolic transition systems $\sts$ in {\sf STS4} and every 
$\mlog_4$-formula~$\varphi$, the symbolic semi-algorithm {\sf ModelCheck} 
terminates on the region algebra $\stt_\sts$ and the input formula~$\varphi$.
}

\mypar
{\bf Corollary 4B} 
{\it 
The $\mlog_4$ and $\dual{\mlog_4}$ model-checking problems are decidable 
for the class {\sf STS4} of symbolic transition systems.
}

\section{Class-5 Symbolic Transition Systems}

We define two states of a transition system to be ``bounded-reach equivalent''
if for every distance~$d$, the same observables can be reached in $d$ or 
fewer transitions.
Class-5 systems are characterized by finite bounded-reach-equivalence 
quotients.
Equivalently, for every observable $\ob$ there is a finite bound $n_\ob$ such 
that all states that can reach $\ob$ can do so in at most $n_\ob$ transitions.
This enables the model checking of all reachability and (by duality) 
invariance properties.
The transition systems in class~5 have also been called 
``well-structured''~\cite{ACJY96}.
Infinite-state examples of class-5 systems are provided by networks of 
rectangular hybrid automata.

\subsection{Finite Characterization: Bounded-distance Targets}

{\bf Definition: Bounded-reach equivalence}
Let $\sts$ be a transition system.
Two states $\state$ and $\astate$ of $\sts$ are 
{\em bounded-reach equivalent}, denoted $\state\cong_5^\sts\astate$, if for
every source-$\state$ trace of $\sts$ with length $n$ and target~$\ob$, 
there is a source-$\astate$ trace of $\sts$ with length at most $n$ and
target~$\ob$, and vice versa.  
The state equivalence $\cong_5$ is called {\em bounded-reach equivalence}.
\qed

\mypar
{\bf Definition: Class {\sf STS5}}
A symbolic transition system $\sts$ belongs to the class {\sf STS5} if the 
bounded-reach-equivalence relation $\cong^5_\sts$ has finite index.
\qed

\begin{figure}[t]
\centering{
\setlength{\unitlength}{2842sp}%
\begingroup\makeatletter\ifx\SetFigFont\undefined%
\gdef\SetFigFont#1#2#3#4#5{%
  \reset@font\fontsize{#1}{#2pt}%
  \fontfamily{#3}\fontseries{#4}\fontshape{#5}%
  \selectfont}%
\fi\endgroup%
\begin{picture}(2708,2377)(143,-1568)
\thinlines
\put(1201,389){\circle{300}}
\put(1171,344){\makebox(0,0)[lb]{\smash{\SetFigFont{9}{10.8}{\rmdefault}{\mddefault}{\updefault}$p$}}}
\put(1201,-511){\circle{300}}
\put(1171,-556){\makebox(0,0)[lb]{\smash{\SetFigFont{9}{10.8}{\rmdefault}{\mddefault}{\updefault}$p$}}}
\put(2101,389){\circle{300}}
\put(2071,344){\makebox(0,0)[lb]{\smash{\SetFigFont{9}{10.8}{\rmdefault}{\mddefault}{\updefault}$p$}}}
\put(2101,-511){\circle{300}}
\put(2071,-556){\makebox(0,0)[lb]{\smash{\SetFigFont{9}{10.8}{\rmdefault}{\mddefault}{\updefault}$p$}}}
\put(2101,-1411){\circle{300}}
\put(2071,-1456){\makebox(0,0)[lb]{\smash{\SetFigFont{9}{10.8}{\rmdefault}{\mddefault}{\updefault}$p$}}}
\put(301,389){\circle{300}}
\put(256,344){\makebox(0,0)[lb]{\smash{\SetFigFont{9}{10.8}{\rmdefault}{\mddefault}{\updefault}$p$}}}
\put(1201,239){\vector( 0,-1){600}}
\put(2101,239){\vector( 0,-1){600}}
\put(2101,-661){\vector( 0,-1){600}}
\put(2851,-511){\makebox(0,0)[lb]{\smash{\SetFigFont{14}{16.8}{\rmdefault}{\mddefault}{\updefault}$. . . .$}}}
\put(151,614){\makebox(0,0)[lb]{\smash{\SetFigFont{9}{10.8}{\rmdefault}{\mddefault}{\updefault}$s_0$}}}
\put(1051,614){\makebox(0,0)[lb]{\smash{\SetFigFont{9}{10.8}{\rmdefault}{\mddefault}{\updefault}$s_1$}}}
\put(1951,614){\makebox(0,0)[lb]{\smash{\SetFigFont{9}{10.8}{\rmdefault}{\mddefault}{\updefault}$s_2$}}}
\end{picture}
}
\caption{Bounded-reach equivalence is coarser than distance equivalence}
\label{sts4sts5}
\end{figure}
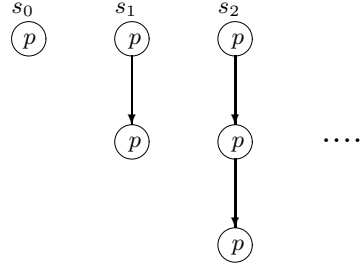

\mypar
Figure~\ref{sts4sts5} shows that bounded-reach equivalence is coarser than 
distance equivalence 
(all states~$\state_i$, for $i\ge 0$, are bounded-reach equivalent, but no 
two of them are distance equivalent).
It follows that the class {\sf STS5} of symbolic transition systems is a 
proper extension of {\sf STS4}.

\subsection{Symbolic State-space Exploration: Predecessor Aggregation}

\begin{figure}[t]
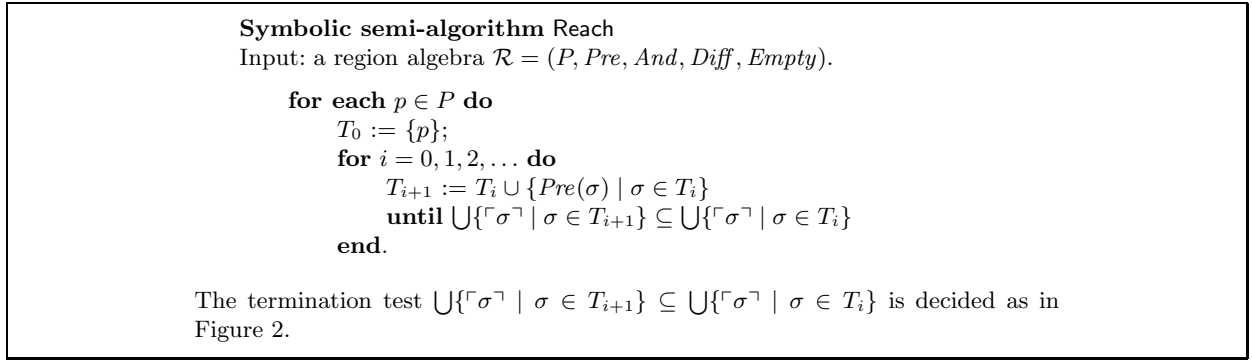

\boxpar{
\begin{verse}
  {\bf Symbolic semi-algorithm} {\sf Reach}\\
  Input: a region algebra $\stt=(\obs,\Pre,\And,\Diff,\Empty)$.\\[5pt]
  \q \FOREACH\ $\ob\in\obs$ \DO\\
  \q\q $T_0$ := $\set{\ob}$;\\
  \q\q \FOR\ $i=0,1,2,\ldots$ \DO\\
  \q\qq $T_{i+1}$ := $T_i\cup\set{\Pre(\reg)\mid\reg\in T_i}$\\
  \q\qq \UNTIL\ $\bigcup\set{\ext{\reg}\mid\reg\in T_{i+1}}\subseteq
      \bigcup\set{\ext{\reg}\mid\reg\in T_i}$\\
  \q\q \END.
\end{verse}
The termination test 
$\bigcup\set{\ext{\reg}\mid\reg\in T_{i+1}}\subseteq
  \bigcup\set{\ext{\reg}\mid\reg\in T_i}$
is decided as in Figure~\ref{modelcheck}.
}
\caption{Predecessor aggregation}
\label{reach}
\end{figure}

The symbolic semi-algorithm {\sf Reach} of Figure~\ref{reach} starts
from the observables and repeatedly applies the $\Pre$ operation, but its
termination criterion is more easily met than the termination criterion
of the semi-algorithm {\sf Closure4}; 
that is, {\sf Reach} may terminate on more inputs than {\sf Closure4}.
Indeed, we shall show that, when the input is the region algebra of a
symbolic transition system $\sts=(\states,\trans,\regs,\ext{\cdot},\obs)$,
then {\sf Reach} terminates iff $\sts$ belongs to the class {\sf STS5}.
Furthermore, upon termination, $\state \cong^\sts_5 \astate$ iff for
each observation $\ob\in\obs$ and each region $\reg\in T_i^\ob$, we have 
$\state\in\ext{\reg}$ iff $\astate\in\ext{\reg}$.

\mypar
An alternative characterization of the class {\sf STS5} can be given using 
well-quasi-orders on states \cite{ACJY96,FinkelSchnoebelen98}.
A {\em quasi-order\/} on a set $A$ is a reflexive and transitive binary 
relation on~$A$.
A {\em well-quasi-order\/} on $A$ is a quasi-order $\preceq$ on $A$ such that 
for every infinite sequence $a_0,a_1,a_2,\ldots$ of elements $a_i\in A$ there 
exist indices $i$ and $j$ with $i<j$ and $a_i\preceq a_j$.
A set $B\subseteq A$ is {\em upward-closed\/} if for all $b\in B$ and 
$a\in A$, if $b\preceq a$, then $a\in B$.
It can be shown that if $\preceq$ is a well-quasi-order on~$A$, then every 
infinite increasing sequence 
$B_0\subseteq B_1\subseteq B_2\subseteq\cdots$ of upward-closed sets 
$B_i\subseteq A$ eventually stabilizes;
that is, there exists an index $i\ge 0$ such that $B_j=B_i$ for all $j\ge i$.

\mypar
{\bf Theorem 5A.}
{\it
For all symbolic transition systems~$\sts$, the following three conditions 
are equivalent:
\begin{verse}
  1. $\sts$ belongs to the class {\sf STS5}.\\
  2. The symbolic semi-algorithm {\sf Reach} terminates on the region 
   algebra~$\stt_\sts$.\\
  3. There is a well-quasi-order $\preceq$ on the states of 
   $\sts$ such that for all observations $\ob$ and all nonnegative 
   integers~$d$, the set $\sem{\EDB{\le d}\ob}_\sts$ is upward-closed.
\end{verse}
}

\mypar
{\bf Proof} 
(2 $\Ra$ 1)
Define $\state\sim^\sts_{\le n}\astate$ if for all observations~$\ob$, for 
every source-$\state$ trace with length $n$ and target~$\ob$, there is a 
source-$\astate$ trace with length at most $n$ and target~$\ob$, and vice 
versa.
Note that $\sim^\sts_{\le n}$ has finite index for all $n\ge 0$.
Suppose that the semi-algorithm {\sf Reach} terminates in at most $i$
iterations for each observation~$\ob$.
Then for all $n\ge i$, the equivalence relation $\sim^\sts_{\le n}$ is equal 
to $\sim^\sts_{\le i}$.
Since $\cong^\sts_5$ is equal to $\bigcap\set{\sim^\sts_{\le n}\mid n\ge 0}$,
it has finite index.

\mypar
(1 $\Ra$ 3)
Define the quasi-order $\state\preceq_5^\sts\astate$ if for all observables 
$\ob$ and all $n\ge 0$, for every source-$\state$ trace with length $n$ and 
target~$\ob$, there is a source-$\astate$ trace with length at most $n$ and
target~$\ob$.
Then each set $\sem{\EDB{\le d}\ob}_\sts$, for an observable $\ob$ and a
nonnegative integer~$d$, is upward-closed with respect to~$\preceq_5^\sts$.
Furthermore, if $\cong^\sts_5$ has finite index, then $\preceq_5^\sts$ is a 
well-quasi-order.
This is because $\state\cong_5^\sts\astate$ implies 
$\state\preceq_5^\sts\astate$:
if there were an infinite sequence $\state_0,\state_1,\state_2,\ldots$ of 
states such that for all $i\ge 0$ and $j<i$, we have 
$\state_j\not\preceq_5^\sts\state_i$, then no two of these states would be 
$\cong_5^\sts$ equivalent.

\mypar
(3 $\Ra$ 2)
This part of the proof follows immediately from the stabilization property of
well-quasi-orders~\cite{ACJY96}.
\qed

\subsection{Decidable Properties: Bounded Reachability}

{\bf Definition: Bounded-reachability logic}
The {\em bounded-reachability logic\/} $\tlog_5$ consists of the 
$\tlog_4$-formulas that are generated by the grammar
  \[\varphi\ ::=\
    \ob \mid
    \varphi\vee\varphi \mid 
    \EDB{\le d}\varphi \mid
    \ED\varphi,
    \]
for constants $\ob\in\consts$ and nonnegative integers~$d$.
The state logic $\dual{\tlog_5}$ consists of the duals of all 
$\tlog_5$-formulas. 
\qed

\mypar
The following facts about bounded-reachability logic and its dual are 
relevant in our context.
Both $\tlog_5$ and $\dual{\tlog_5}$ admit abstraction, and the state 
equivalence induced by both $\tlog_5$ and $\dual{\tlog_5}$ is~$\cong_5$ 
(bounded-reach equivalence).
It follows that the conjunction-free temporal logic $\tlog_4$ is more 
expressive than~$\tlog_5$, and $\dual{\tlog_4}$ is more expressive 
than~$\dual{\tlog_5}$.
For example, the property that an observable can be reached in exactly $d$ 
transitions can be expressed in $\tlog_4$ but not in~$\tlog_5$.
Since $\tlog_5$ admits abstraction, and for {\sf STS5} systems the induced 
quotient can be constructed using the symbolic semi-algorithm {\sf Reach},
we have the following theorem.

\mypar
{\bf Theorem 5B} 
{\it 
The $\tlog_5$ and $\dual{\tlog_5}$ model-checking problems are decidable for 
the class {\sf STS5} of symbolic transition systems.
}

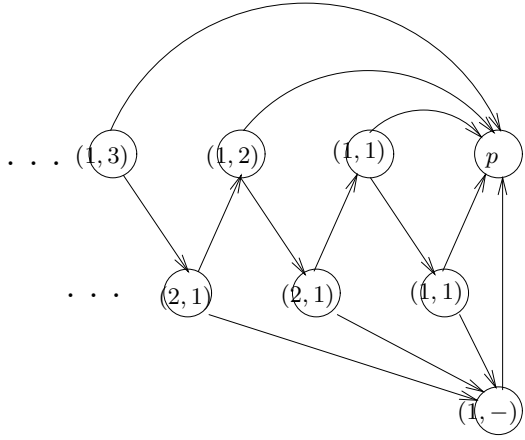
\begin{figure}[t]
\centering{
\setlength{\unitlength}{0.00075000in}
\begingroup\makeatletter\ifx\SetFigFont\undefined%
\gdef\SetFigFont#1#2#3#4#5{%
  \reset@font\fontsize{#1}{#2pt}%
  \fontfamily{#3}\fontseries{#4}\fontshape{#5}%
  \selectfont}%
\fi\endgroup%
{
\begin{picture}(3628,3057)(0,-10)
\put(1650,1975){\ellipse{336}{336}}
\put(2550,1975){\ellipse{336}{336}}
\put(3450,1975){\ellipse{336}{336}}
\put(2100.000,1542.542){\arc{2980.054}{3.5542}{5.8706}}
\path(3385.773,2234.990)(3465.000,2140.000)(3439.795,2261.098)
\put(2536.637,1446.493){\arc{2227.736}{3.8137}{5.6282}}
\path(3318.825,2196.160)(3420.000,2125.000)(3364.215,2235.400)
\put(2926.314,1754.175){\arc{1056.644}{3.9196}{5.5820}}
\path(3221.963,2155.233)(3330.000,2095.000)(3263.018,2198.988)
\put(2175,1000){\ellipse{336}{336}}
\put(3075,1000){\ellipse{336}{336}}
\put(1275,1000){\ellipse{336}{336}}
\put(3450,175){\ellipse{336}{336}}
\path(1350,1150)(1650,1825)
\path(1628.678,1703.158)(1650.000,1825.000)(1573.849,1727.527)
\path(2161,1156)(2461,1831)
\path(2439.678,1709.158)(2461.000,1831.000)(2384.849,1733.527)
\path(3061,1156)(3361,1831)
\path(3339.678,1709.158)(3361.000,1831.000)(3284.849,1733.527)
\path(825,1825)(1280,1150)
\path(1188.050,1232.736)(1280.000,1150.000)(1237.803,1266.273)
\put(750,1975){\ellipse{336}{336}}
\path(1425,850)(3300,250)
\path(3176.566,258.000)(3300.000,250.000)(3194.852,315.146)
\put(420,970){\makebox(0,0)[lb]{\smash{{{\SetFigFont{14}{16.8}{\rmdefault}{\mddefault}{\updefault}. . .}}}}}
\path(2325,850)(3345,310)
\path(3224.909,339.633)(3345.000,310.000)(3252.982,392.660)
\path(3180,850)(3435,340)
\path(3354.502,433.915)(3435.000,340.000)(3408.167,460.748)
\path(1649,1805)(2099,1130)
\path(2007.474,1213.205)(2099.000,1130.000)(2057.397,1246.487)
\path(2555,1809)(3005,1134)
\path(2913.474,1217.205)(3005.000,1134.000)(2963.397,1250.487)
\path(3480,325)(3480,1810)
\path(3510.000,1690.000)(3480.000,1810.000)(3450.000,1690.000)
\put(0,1900){\makebox(0,0)[lb]{\smash{{{\SetFigFont{14}{16.8}{\rmdefault}{\mddefault}{\updefault}. . .}}}}}
\put(3375,115){\makebox(0,0)[b]{\smash{{{\SetFigFont{9}{10.8}{\familydefault}{\mddefault}{\updefault}$(1,-)$}}}}}
\put(3405,1915){\makebox(0,0)[b]{\smash{{{\SetFigFont{9}{10.8}{\familydefault}{\mddefault}{\updefault}$p$}}}}}
\put(2475,1945){\makebox(0,0)[b]{\smash{{{\SetFigFont{9}{10.8}{\familydefault}{\mddefault}{\updefault}$(1,1)$}}}}}
\put(675,1915){\makebox(0,0)[b]{\smash{{{\SetFigFont{9}{10.8}{\familydefault}{\mddefault}{\updefault}$(1,3)$}}}}}
\put(1590,1915){\makebox(0,0)[b]{\smash{{{\SetFigFont{9}{10.8}{\familydefault}{\mddefault}{\updefault}$(1,2)$}}}}}
\put(3015,955){\makebox(0,0)[b]{\smash{{{\SetFigFont{9}{10.8}{\familydefault}{\mddefault}{\updefault}$(1,1)$}}}}}
\put(2115,940){\makebox(0,0)[b]{\smash{{{\SetFigFont{9}{10.8}{\familydefault}{\mddefault}{\updefault}$(2,1)$}}}}}
\put(1260,925){\makebox(0,0)[b]{\smash{{{\SetFigFont{9}{10.8}{\familydefault}{\mddefault}{\updefault}$(2,1)$}}}}}
\end{picture}
}
}
\label{OOp}
\caption{An {\sf STS5} system on which $\mlog_4$ does not terminate}
\end{figure}

\mypar
A direct symbolic model-checking semi-algorithm for $\tlog_5$ and, indeed,
$\tlog_4$ is easily derived from the semi-algorithm {\sf Reach}. 
Then, if {\sf Reach} terminates, so does model checking for all
$\tlog_4$-formulas, including unbounded $\ED$ properties.
The extension to $\tlog_4$ is possible, because $\EN$ properties pose no 
threat to termination.
However this is not true for $\mlog_4$:
Figure~\ref{OOp} shows a symbolic transition system in the class
{\sf STS5} for which the naive evaluation of the formula 
$(\mu x: p \vee \EN\EN x)$ does not terminate.
We now show that this is not suprising as $\mlog_4$ is undecidable on
{\sf STS5} systems.
To establish this result, we proceed as follows: given a two-counter machine 
$M=\langle \{ b_1,\cdots,b_m\},C,D \rangle$, we define a symbolic 
transition system $\sts_M$ that belongs to the class {\sf STS5} and
that encodes the computations of $M$ using $\Pre^2$.
On such a structure we prove that the formula 
$(\mu x: {\sf Final} \vee \EN\EN x)$ characterizes exactly the set of 
configurations of the two-counter machine that can reach a final location.
This will establish the undecidability of $\mlog_4$ on {\sf STS5} systems.

Without lost of generality, we make the following hypothesis on the 
two-counter machine $M$: there is only one initial location and only
one final location in $M$, we denote them $b_0$ and $b_m$
respectively. Furthermore,
the initial location of $M$ is never reached after the first instruction.
A {\em configuration} of $M$ is a triple $\gamma=\langle i,c,d \rangle$, where
$i$ is the program counter indicating the current instruction, and $c$ and
$d$ are the values of the counters $C$ and $D$.
A {\em computation} of $M$ is a finite or infinite sequence 
$\sigma=\gamma_0 \gamma_1 \dots$ of configurations such that
for every $\gamma_{i+1}$ is a $M$-successor of $\gamma_i$. 
In the sequel, we write $(\gamma_i,\gamma_{i+1})\in R_M$ to denote that
$\gamma_{i+1}$ is a $M$-successor of $\gamma_i$.
We say that a computation $\sigma$ is {\em initial} if 
$\gamma_0=\langle 0,0,0 \rangle$, that is the first instruction is the 
initial instruction and the two counters have the value $0$.
We say that a computation $\sigma$ is {\em final} if $\sigma$ 
is finite and its last configuration contains the stop instruction.
The {\em halting problem} for a two-counter machine $M$ is to decide whether
or not the execution of $M$ has at least one initial computation 
that ends in a stop instruction.
The problem of deciding if a two-counter machine has a halting computation
is undecidable \cite{HopcroftUllman79}.

We define the transition system $\sts_M$ that encodes the computations 
of $M$ using $\Pre^2$ as follows. 
\begin{itemize}
   \item The states of the transition system are pairs $(\gamma,i)$ 
	where $\gamma$ is a configuration of $M$ and $i \in \{1,2\}$. 
	We call $(\gamma,1)$ the copy-1 of configuration $\gamma$, and 
	$(\gamma,2)$ the copy-2 of configuration $\gamma$. 
	Formally the set of states $Q$ is the union $I \cup B \cup F$, where :
	(i) $I=\{ (\langle 0,0,0 \rangle,1) \}$, that is, the singleton 
	containing the copy-1 of the initial configuration of $M$;
	(ii) $B=\{ (\langle 0,0,0 \rangle, 2) \}
		\cup \{ (\langle b,c,d \rangle,i) \mid 
				b\not= 0 
				\land b \not=m
				\land c,d \geq 0 \}$, that is, the set
		containing the copy-2 of the initial configuration of $M$
		and two copies of each configuration of $M$ which is 
		not initial and not final;
	(iii) $F=\{ (\langle b,c,d \rangle,1) 
				\mid b=m
					\land c,d \geq 0 \}$, that is
		the copy-1 of each final configurations of $M$.

  \item The transition relation $\trans$ is defined as follows:
	for every $(\gamma_1,i_1), (\gamma_2,i_2) \in Q$, we have that
	$(\gamma_2,i_2) \in \trans(\gamma_1,i_1)$ if and only if
	one of the following conditions is satisfied :
	(i) $(\gamma_1,i_1) \in I \cup B \land i_1=1 \land
		(\gamma_2,i_2) \in F$, that is every copy-1 of a configuration
		which is not final is linked to every final configuration;
	(ii) $\gamma_1=\gamma_2 \land i_1=1 \land i_2=2 \land 
		(\gamma_1,i_1) \in B \cup I$, that is every copy-1
		of a configuration $\gamma$ is linked to the copy-2 of
		$\gamma$;
	(iii) $(\gamma_1,i_1) \in B \land i_1=2 \land (\gamma_2,i_2) 
		\in B \cup F \land i_2=1 \land (\gamma_1,\gamma_2) \in R_M$,
		that is the copy-2 of a configuration $\gamma_1$
		is linked to the copy-1 of a configuration $\gamma_2$
		if $\gamma_2$ is a $M$-successor of $\gamma_1$.
  \item The set of regions $R$ is the set of sets of states definable
		by Presburger formulas.
  \item The set of propositions $P$ is $\{ {\sf Init},{\sf Between},
		{\sf Final}\}$, with the following extension function:
		(i) $\ext{{\sf Init}}=I$, 
		(ii) $\ext{{\sf Between}}=B$, and
		(iii) $\ext{{\sf Final}}=F$.
\end{itemize}
We now establish three properties of the symbolic transition system $\sts_M$.

\mypar
{\bf Lemma 5A}
{\it
Presburger formulas form a region algebra for the transition system $\sts_M$.
}

\mypar
{\bf Proof.}
This algebra is trivially closed under all boolean operations, furthermore
the problems of satisfiability and of membership for Presburger formulas are 
decidable.
So, it remains us to show that the set of states satisfying the propositions 
are expressible as Presburger formula and for all regions $R$, $\Pre(R)$ 
is expressible by a Presburger formula.
Let us consider the proposition ${\sf Between}$, 
the set of states of $\sts_M$ that
satisfy ${\sf Between}$ is expressed by the following Presburger formula:
$0<i<m \land (( c \geq 0 \land d \geq 0 \land (copy=1 \lor copy=2)) \lor
		(i=0 \land copy=2 \land c=0 \land d=0))$. 
The other propositions are left to the reader.
Let us now show that the region algebra is closed under $\Pre$.
We show how to construct the formula $\Phi$ that represent $\Pre(R_{\Psi})$,
where $R_{\Psi}$ is the set of states defined by the Presburger formula
$\Psi$ with free variable $i',c',d',copy'$.
By definition of $\trans$, we have to consider three cases.
We treat the third one, the two first are trivial and left to the reader.
The final formula is obtained by taking the disjunction of the three formulas.
To construct the formula for the third case, we proceed as follows.
For each instruction $j$ of the two-counter machine, we construct 
a Presburger formula. We treat the case where the instruction $j$ is of the
form $i_1~:~c:=c+1\rightarrow i_2$. The corresponding Presburger 
formula is:
$\exists i',c',d',copy' \cdot \Psi(i',c',d',copy') \land i=i_1 \land i=i_2
	\land c=c'-1 \land d=d' \land copy=2 \land copy'=1$.
\qed

\mypar
{\bf Lemma 5B}
{\it
The transition system $\sts_M$ is in the class {\sf STS5}.
}

\mypar
{\bf Proof.}
We show that for every proposition $p \in P$, the iteration of $\Pre$
terminates:
  \begin{itemize}
	\item $p \equiv {\sf Init}$. Trivially, $\Pre^*(\ext{{\sf Init}})=I$
		as $\ext{{\sf Init}}=I=\{ (\langle 0,0,0 \rangle,1) \}$
		and $(\langle 0,0,0 \rangle,1)$ has no predecessors by
		definition of $\trans$;
	\item $p \equiv {\sf Between}$.
		We have $\Pre^*(\ext{{\sf Between}})=I \cup B=
		Pre^{\leq 1}(B)$, in fact the copy-1 of the 
		initial configuration of $M$ is reached after one
		iteration, no other states can be added (the states
		of $F$ has no outgoing edges).
	\item $p \equiv {\sf Final}$.
		We have $\Pre^*(\ext{{\sf Final}})=Pre^{\leq 2}
		(\ext{{\sf Final}})$, in fact $\ext{{\sf Final}}=F$,
		$\Pre(F)=I \cup \{ (\gamma_1,i_1) \in B \mid
				i_1=1 
				\lor 
				( \exists (\gamma_2,1) \in F 
				   \land (\gamma_1,\gamma_2) \in R_M) \}$,
		and $\Pre(\Pre(F)) \supseteq \{ (\gamma_1,2) \mid
			\exists (\gamma_2,1) \in B \land (\gamma_2,1) \in
					\trans((\gamma_1,2)) \}$,
		thus $\Pre^{\leq 2}(\ext{{\sf Final}})$ contains
		every states of $Q$ that is either final or 
		has at least one outgoing edge, and
		thus no other state can be added.
  \end{itemize}
\qed

\mypar
{\bf Lemma 5C}
{\it
For every $(\gamma_a,1) \in Q$, $(\gamma_a,1) \in (Pre^2)^n(F)$ if
and only if there exists a computation $\sigma=\gamma_0 \gamma_1 \dots
\gamma_n$ of $M$ such that $\gamma_a=\gamma_0$ and $\gamma_n$ is a final
configuration.
}

\mypar
{\bf Proof.}
Let us first establish the left to right direction.
We reason by induction on $i$.
Base case: $i=0$. As $(Pre^2)^0(F)=F$, this is trivial.
Induction case: $i=k>0$. Let us consider 
$(\gamma_a,1) \in (Pre^2)((Pre^2)^{k-1}(F))$.
By construction of $\sts_M$, we know that
$B \cap \trans{(\gamma_a,1)}=\{ (\gamma_a,2) \}$ and 
$\trans{(\gamma_a,2)}=\{(\gamma_c,1) \mid (\gamma_a,\gamma_c) \in R_M \}$.
By hypothesis, $\trans{(\gamma_a,2)} \cap (Pre^2)^{k-1}(F)$ is non-empty.
Consider $(\gamma_c,1) \in \trans{(\gamma_a,2)} \cap (Pre^2)^{k-1}(F)$,
by induction hypothesis, there exists a final $M$-computation
$\sigma'=\gamma_0 \gamma_1 \dots \gamma_{k-1}$.
We construct $\sigma=\gamma_a \cdot \sigma'$ which is a final $M$-computation
that goes from $\gamma_a$ to a final configuration of $M$.

Let us now establish the right to left implication.
We show that if $\sigma=\gamma_0 \gamma_1 \dots \gamma_n$ is a final 
$M$-computation then $(\gamma_1,1) \in (Pre^2)^{n}(F)$.
We reason by induction on the value of $n$.
Base case : $n=0$, that is $\sigma=\gamma_0$. In this case,
$\gamma_0$ is a final configuration and $(\gamma_0,1) \in F$ and
trivially, $(\gamma_0,1) \in (Pre^2)^0(F)$.
Induction case: $n=k>0$.
Let us consider the final $M$-computation 
$\sigma=\gamma_0 \gamma_1 \dots \gamma_{k}$.
By definition $\gamma_1 \dots \gamma_{k}$ is a final $M$-computation and
by induction hypothesis $(\gamma_1,1) \in (Pre^2)^{k-1}(F)$.
Let us show that $(\gamma_0,1) \in Pre^2(\{(\gamma_1,1)\})$ holds.
We know that $(\gamma_0,\gamma_1) \in R_M$ as $\sigma$ is a $M$-computation,
and by definition of $\trans{}$, we have 
$(\gamma_1,1) \in \trans{(\gamma_0,2)}$, and as 
$(\gamma_0,2) \in \trans{(\gamma_0,1)}$, we have
$(\gamma_0,1) \in Pre^2(\{(\gamma_1,1)\})$.
It follows that $(\gamma_0,1) \in (Pre^2)^k(F)$.   
\qed

From the above lemmas, it follows that
the $\mlog_4$ formula $(\mu x: {\sf Final} \vee \EN\EN x)$ expresses on 
$\sts_M$ exactly the set of configurations of $M$ that can reach a final
location of $M$.
The undecidability of model-checking $\mlog_4$ on the class {\sf STS5}
follows as a consequence.

\mypar
{\bf Theorem 5B-Undecidability} 
{\it 
The $\mlog_4$ and $\dual{\mlog_4}$ model-checking problems are undecidable for 
the class {\sf STS5} of symbolic transition systems.
}

\subsection{Example: Networks of Rectangular Hybrid Automata}

A {\em network of timed automata\/}~\cite{AbdullaJonsson98} consists of a
finite state controller and an arbitrarily large set of identical 1D timed
automata.
The continuous evolution of the system increases the values of all 
variables.
The discrete transitions of the system are specified by a set of
synchronization rules.
We generalize the definition to rectangular automata.
Formally, a {\em network of rectangular automata\/} is a triple 
$(C, \hyb, R)$, where $C$ is a finite set of controller locations, $\hyb$ is 
a 1D rectangular automaton, and $R$ is a finite set of rules of the form 
$r=(\langle c, c'\rangle, e_1, \ldots, e_n)$, where $c,c'\in C$ and 
$e_1,\ldots,e_n$ are jumps of~$\hyb$.
The rule $r$ is enabled if the controller state is $c$ and there are $n$ 
rectangular automata $\hyb_1,\ldots,\hyb_n$ whose states are such that the 
jumps $e_1,\ldots,e_n$, respectively, can be performed.
The rule $r$ is executed by simultaneously changing the controller state to 
$c'$ and the state of each~$\hyb_i$, for $1\le i\le n$, according to the 
jump~$e_i$.
The following result is proved in \cite{AbdullaJonsson98} for networks of 
timed automata.
The proof can be extended to rectangular automata using the observation
that every rectangular automaton is simulated by an appropriate timed 
automaton~\cite{STOC95j}.

\mypar
{\bf Theorem 5C} 
{\it 
The networks of rectangular automata belong to the class {\sf STS5}.
There is a network of timed automata that does not belong to {\sf STS4}.
}

\section{General Symbolic Transition Systems}

For studying reachability questions on symbolic transition systems, it is
natural to consider the following fragment of bounded-reachability logic.

\mypar
{\bf Definition: Reachability logic}
The {\em reachability logic\/}  $\tlog_6$ consists of the $\tlog_5$-formulas 
that are generated by the grammar
  \[\varphi\ ::=\
    \ob \mid
    \varphi\vee\varphi \mid 
    \ED\varphi,
    \]
for constants $\ob\in\consts$.
\qed

\begin{figure}[t]
\centering{
\setlength{\unitlength}{2842sp}%
\begingroup\makeatletter\ifx\SetFigFont\undefined%
\gdef\SetFigFont#1#2#3#4#5{%
  \reset@font\fontsize{#1}{#2pt}%
  \fontfamily{#3}\fontseries{#4}\fontshape{#5}%
  \selectfont}%
\fi\endgroup%
\begin{picture}(2870,562)(1043,82)
\thinlines
\put(2101,239){\circle{300}}
\put(3001,239){\circle{300}}
\put(1201,239){\circle{300}}
\put(1951,239){\vector(-1, 0){600}}
\put(2851,239){\vector(-1, 0){600}}
\multiput(3901,239)(-166.66667,0.00000){5}{\line(-1, 0){ 83.333}}
\put(3151,239){\vector(-1, 0){0}}
\put(2926,464){\makebox(0,0)[lb]{\smash{\SetFigFont{9}{10.8}{\rmdefault}{\mddefault}{\updefault}$s_1$}}}
\put(2026,464){\makebox(0,0)[lb]{\smash{\SetFigFont{9}{10.8}{\rmdefault}{\mddefault}{\updefault}$s_0$}}}
\put(2041,179){\makebox(0,0)[lb]{\smash{\SetFigFont{9}{10.8}{\rmdefault}{\mddefault}{\updefault}$p$}}}
\put(2941,179){\makebox(0,0)[lb]{\smash{\SetFigFont{9}{10.8}{\rmdefault}{\mddefault}{\updefault}$p$}}}
\put(1141,179){\makebox(0,0)[lb]{\smash{\SetFigFont{9}{10.8}{\rmdefault}{\mddefault}{\updefault}$q$}}}
\end{picture}
}
\caption{Reach equivalence is coarser than bounded-reach equivalence}
\label{sts5sts6}
\end{figure}
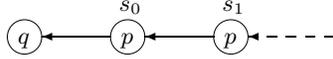

\mypar
The reachability logic $\tlog_6$ is less expressive than the
bounded-reachability logic~$\tlog_5$, because it induces the following state
equivalence, $\cong_6$, which is coarser than bounded-reach equivalence 
(see Figure~\ref{sts5sts6}: all states~$\state_i$, for $i\ge 0$, are reach 
equivalent, but no two of them are bounded-reach-equivalent).

\mypar
{\bf Definition: Reach equivalence}
Let $\sts$ be a transition system.
Two states $\state$ and $\astate$ of $\sts$ are {\em reach equivalent},
denoted $\state\cong_6^\sts\astate$, if for every source-$\state$ trace of 
$\sts$ with target~$\ob$, there is a source-$\astate$ trace of $\sts$ with 
target~$\ob$, and vice versa.
The state equivalence $\cong_6$ is called {\em reach equivalence}.
\qed

\mypar
For every symbolic transition system $\stt$ with $k$ observables, the
reach-equivalence relation $\cong^\stt_6$ has at most $2^k$ equivalence
classes and, therefore, finite index.
Since the reachability problem is undecidable for many kinds of symbolic
transition systems 
(including Turing machines and polyhedral hybrid automata~\cite{DES94j}),
it follows that there cannot be a general algorithm for computing the 
reach-equivalence quotient of symbolic transition systems.

\newcommand{\etalchar}[1]{$^{#1}$}

\end{document}